\documentclass[a4paper,fleqn]{mnras}
\usepackage{threeparttable}
\usepackage{multicol}
\usepackage{aas_macros,graphicx,times,multirow,amsmath,amssymb,color,longtable}
\usepackage[T1]{fontenc}

\title[Multi-band study of the field of 3FGL\,J0838.8$-$2829]{Multi-band study of RX\,J0838$-$2827 and XMM J083850.4$-$282759: a new asynchronous magnetic cataclysmic variable and a candidate transitional millisecond pulsar}

\author[N. Rea et al.]{N.~Rea,$^{1,2}$\thanks{E-mail: rea@ice.csic.es} F.~Coti Zelati,$^{1,2,3,4}$  P.~Esposito,$^{2}$ P. D'Avanzo,$^{4}$ D. de~Martino,$^{5}$ G. L. Israel,$^{6}$ \newauthor D. F. Torres,$^{1,7}$ S. Campana,$^{4}$ T. M. Belloni,$^{4}$ A. Papitto,$^{6}$ N. Masetti,$^{8,9}$, L. Carrasco,$^{10}$  \newauthor A. Possenti,$^{11}$ M. Wieringa,$^{12}$ E. De O\~na Wilhelmi,$^{1}$ J. Li,$^{1}$ E. Bozzo,$^{13}$ C. Ferrigno,$^{13}$ \newauthor M. Linares,$^{14,15,16}$ T. M. Tauris,$^{17,18}$ M. Hernanz,$^{1}$ I. Ribas,$^{1}$ M. Monelli$^{14,15}$ \newauthor   A. Borghese,$^{2}$ M. C. Baglio,$^{3,4}$ J. Casares$^{14,15,19}$ 
  \smallskip\\
$^{1}$ Institute of Space Sciences (IEEC--CSIC), Carrer de Can Magrans s/n, E-08193 Barcelona, Spain\\
$^{2}$ Anton Pannekoek Institute for Astronomy, University of Amsterdam, Postbus 94249,  NL-1090-GE Amsterdam, The Netherlands\\
$^{3}$ Universit\`a dell' Insubria, via Valleggio 11, I-22100 Como, Italy\\
$^{4}$ INAF -- Osservatorio Astronomico di Brera, via Bianchi 46, I-23807 Merate (LC), Italy\\
$^{5}$ INAF -- Osservatorio Astronomico di Capodimonte, Salita Moiariello 16, I-80131 Napoli, Italy \\
$^{6}$ INAF -- Osservatorio Astronomico di Roma, via Frascati 33, I-00078, Monteporzio Catone (RM), Italy\\ 
$^{7}$ Instituci\'o Catalana de Recerca i Estudis Avancats (ICREA), E-08010 Barcelona, Spain \\
$^{8}$ INAF -- Istituto di Astrofisica Spaziale e Fisica Cosmica di Bologna, via Gobetti 101, I-40129, Bologna, Italy\\ 
$^{9}$ Departamento de Ciencias Fisicas, Universidad Andr\'es Bello, Fern\'andez Concha 700, Las Condes, Santiago, Chile \\
$^{10}$ Instituto Nacional de Astrofisica Optica y Electronica (INAOE), Apartado Postal 51 y 216, 72000 Puebla, Mexico \\ 
$^{11}$ INAF -- Osservatorio Astronomico di Cagliari, Via della Scienza 5, I-09047, Selargius (CA), Italy \\
$^{12}$ Australia Telescope National Facility, CSIRO Astronomy and Space Science, PO box 76, Epping, NSW 1710, Australia\\
$^{13}$ ISDC, University of Geneva, chemin d'Écogia, 16 1290, Versoix, Switzerland \\
$^{14}$  Instituto de Astrof{\'i}sica de Canarias, c/ V{\'i}a L{\'a}ctea s/n, E-38205 La Laguna, Tenerife, Spain \\
$^{15}$Universidad de La Laguna, Dept. Astrof{\'i}sica, E-38206 La Laguna, Tenerife, Spain \\
$^{16}$ Departament de F{\'i}sica, EEBE, Universitat Polit{\`e}cnica de Catalunya, c/ Eduard Maristany 10, 08019 Barcelona, Spain \\
$^{17}$  Max-Planck-Institut f\"ur Radioastronomie, Auf dem H\"ugel 69, 53121, Bonn, Germany\\
$^{18}$ Argelander-Institute f\"ur Astronomie, Universit\"at Bonn, Auf dem H\"ugel 71, 53121, Bonn, Germany \\
$^{19}$  Department of Physics, Astrophysics, University of Oxford, Denys Wilkinson Building, Keble Road, Oxford OX1 3RH, UK 
}
\date{\today}
\pagerange{\pageref{firstpage}--\pageref{lastpage}} 
\pubyear{2016}

\def\ltsima{$\; \buildrel < \over \sim \;$}
\def\lsim{\lower.5ex\hbox{\ltsima}}
\def\gtsima{$\; \buildrel $\geq$ \over \sim \;$}
\def\gsim{\lower.5ex\hbox{\gtsima}}
\newcommand{\be}{\begin{equation}}
\newcommand{\en}{\end{equation}}

\def\deg {$^\circ$}
\def\nh{\hbox{$N_{\rm H}$}}
\def\flux {\mbox{erg cm$^{-2}$ s$^{-1}$}}
\def\lum {\mbox{erg s$^{-1}$}}
\def\rx{\mbox{RX\,J0838$-$2827}} 
\def\tmsp{\mbox{XMM\,J083850.4$-$282759}} 
\def\thirdxmm{\mbox{XMM\,J083842.7$-$283832}}
\def\fermirx{\mbox{3FGL\,J0838.8$-$2829}} 

\newcommand{\xmm}{{\em XMM--Newton}}

\newcommand{\swift}{{\em Swift}}
\newcommand{\fermi}{{\em Fermi}}

\begin{document}

\label{firstpage}
\pagerange{\pageref{firstpage}--\pageref{lastpage}}
\maketitle

\begin{abstract}

In search for the counterpart to the \fermi-LAT source 3FGL\,J0838.8$-$2829, we performed a multi-wavelength campaign, in the X-ray band with \swift\, and \xmm, performed infrared, optical (with OAGH, ESO-NTT and IAC80) and radio (ATCA) observations, as well as analysed archival hard X-ray data taken by {\it INTEGRAL}. We report on three X-ray sources consistent with the position of the \fermi-LAT source. We confirm the identification of the brightest object, \rx, as a magnetic cataclysmic variable, that we recognize as an asynchronous system (not associated with the \fermi-LAT source). \rx\, is extremely variable in the X-ray and optical bands, and timing analysis reveals the presence of several periodicities modulating its X-ray and optical emission. The most evident modulations are interpreted as due to the binary system orbital period of $\sim$1.64\,h and the white dwarf spin period of $\sim$1.47\,h. A  strong flux modulation at $\sim$15\,h is observed at all energy bands, consistent with the beat frequency between spin and orbital periods. Optical spectra show prominent H$\beta$, HeI and HeII emission lines Doppler-modulated at the orbital period and at the beat period. Therefore, \rx\, accretes through a diskless configuration and could be either a strongly asynchronous polar or a rare example of a pre-polar system in its way to reach synchronism. Among the other two X-ray sources, \tmsp\ showed a variable X-ray emission, with a powerful flare lasting $\sim$600\,s, similar to what is observed in transitional millisecond pulsars during the sub-luminous disc state: that would possibly associate this source with the \fermi-LAT source.

\end{abstract}
\begin{keywords}
accretion, accretion discs -- novae, cataclysmic variables -- pulsars: general -- white dwarfs -- X-rays: individual: \rx -- X-rays: individual: \tmsp
\end{keywords}

\section{Introduction}
\subsection{Cataclysmic variables}

Cataclysmic variables (hereafter CVs) are close binary systems hosting a white dwarf (WD) and a low-mass main sequence star typically filling its Roche lobe and transferring material toward the compact object (see Warner 2003 for a review). About 20--25 per cent of the known CVs harbour WDs with magnetic fields in the 10$^5$--10$^8$~G range, and are known as magnetic CVs (mCVs; Ferrario, de Martino \& G\"{a}nsicke 2015). The mCVs can be divided into two main groups: the intermediate polars (IPs) and the polars. The former are characterised by asynchronously rotating WDs ($P_{\mathrm{spin}} \ll P_{\mathrm{orb}}$) with spin periods ranging from few hundreds to a thousands of seconds, and orbital periods from one to tens of hours; typically, they accrete via a truncated accretion disc. Due to the lack of synchronisation and of optical/nIR circular polarisation, the WD is believed to possess moderately low magnetic fields ($\leq$$10^7$~G). The Polars instead consist of orbitally-locked rotating WDs ($P_{\mathrm{spin}} = P_{\mathrm{orb}}\sim$~hrs) and are circularly polarised in the optical/nIR  (see Cropper 1990). Their magnetic field as measured from cyclotron emission ranges between $\sim$10--300\,MG, preventing the formation of an accretion disc; thus, accretion proceeds directly from the donor star  onto the WD magnetic poles through a stream.  

A handful of polars were found to be slightly desynchronised  ($\leq$2\%), namely V1500\,Cyg, BY\,Cam, V1432 Aql and CD\,Ind (Campbell \& Schwope 1999). Another candidate, SWIFT\,J2319.4$+$2619, was found recently (Shafter et al. 2008).  Since V1500 Cyg exploded as a nova in 1975, these asynchronous polars are believed to have undergone a nova explosion in the past and they will return to synchronism in hundreds of years (Boyd et al. 2014; Harrison \& Campbell 2016). However, only V1500 Cyg is known as a nova, and searches for nova shells resulted in no detections (Pagnotta \& Zurek 2016). 

Recently, two systems were found to be desynchronised at a much higher level (i.e. $\gg$2\%), namely Paloma (Schwarz et al. 2007) and IGR\,J19552$+$0044 (Bernardini et al. 2013).  These systems might be either regarded as IPs with a very low degree of asynchronism or polars with a large degree of asynchronism. In the former case these systems are not expected to reach synchronism,  while in the latter case they may represent the true polar progenitors (see Norton et al. 2004).  Despite their small number, asynchronous polars represent excellent laboratories to study the physics of mass transfer in the case of a varying magnetic field geometry, especially because of the continuous changes of the orientation between the WD magnetic field axis and the accretion flow.\\

\subsection{Transitional millisecond pulsars}
The short spin periods of binary millisecond pulsars are the outcome of the accretion onto the neutron star (NS) of the mass transferred by a low mass late type companion star, through an accretion disc (Alpar et al. 1982). After a Gyr-long mass accretion phase during which the binary system shines as a bright low mass X-ray binary (NS-LMXB), the mass transfer rate declines and allows the activation of a---now rapidly spinning---radio/gamma-ray millisecond pulsar (MSP), powered by the rotation of its magnetic field. The tight link existing between radio MSPs and NS-LMXBs has been recently demonstrated by the discovery of three transitional millisecond pulsars (PSR J1023$+$0038, Archibald et al. 2009; IGR J18245$-$2452 in the globular cluster M28, Papitto et al. 2013; XSS J12270$-$4859, de Martino et al. 2014, Bassa et al. 2014). 

These sources have been observed to switch between accretion and rotation-powered emission on timescales ranging from a couple of weeks to months, showing that such state transitions may take place on time scales compatible with those of the variations of the mass accretion rate onto the NS.  At high mass inflow rates, the radio pulsar is shut-off and the system is bright in the X-rays ($L_\mathrm{X} > 10^{36}$\,\lum). At low mass inflow rates, the magnetosphere expands up to the light cylinder activating the radio pulsar, the disc disappears and the system is faint in the X-rays ($L_\mathrm{X}\sim10^{32}$ \lum). 

Surprisingly, in addition to the X-ray outbursts accreting state and the radio pulsar state, the three known transitional MSPs have been observed into another extremely peculiar {\em sub-luminous} disc state with $L_\mathrm{X}\sim10^{33}$ \lum\ (see also Linares 2014). During these states, both PSR J1023$+$0038 (Archibald et al. 2015) and XSS J1227$-$4859 (Papitto et al. 2015) showed that part of the disc material is accreted onto the NS surface. These sub-luminous states are accompanied by X-ray flaring activity, a sizeable gamma-ray emission and a flat-radio spectrum that are typical jet signatures in accreting compact objects, suggesting that large mass outflows could be launched by the fast rotating propellering magnetosphere of these pulsars (Papitto et al. 2014; Papitto \& Torres 2015; but see also Takata et al. (2014) for an alternative modelling based on a pulsar wind/disk shock).

\subsection{The ROSAT source: 1RXS\,J083842.1$-$282723}

1RXS\,J083842.1$-$282723 (\rx\, hereafter) was discovered by ROSAT, and tentatively classified as a CV by Masetti et al. 
(2013) based on Balmer and helium emission lines observed in its optical spectrum. Since two of the three transitional MSPs were formerly misclassified as CVs because of their CV-like optical spectra, searches within the error circles of newly discovered but still unidentified \fermi-LAT sources are being carried out despite early tentative classifications based on optical spectroscopy. In particular, the 
spatial coincidence with the unidentified gamma-ray source 3FGL\,J0838.8$-$2829 in the \fermi\ Large Area Telescope (LAT) 4-year point source catalog (Acero et al. 2015) and the variability of the X-ray emission detected on a time scale of a few hours in archival \swift\ observations, made the nature of the source rather controversial.

We report in this paper on a large unprecedented multi-band campaign of the field of \fermirx\ in the radio, infrared, optical and X-ray bands. In Sections\,\ref{Xobs}, \ref{opt}, and \ref{radio}, we report on the details of the observations of the field of \fermirx . In Section\,\ref{analysis}, data analysis and results on \rx\, are collected. In Section\,\ref{others}, we report on the analysis of the other X-ray sources associated with \fermirx\, and in particular on the flaring source \tmsp, which might be a new transitional millisecond pulsar possibly associated with the gamma-ray emission. Discussion follows in Section\,\ref{discussion}. 

\section{X-ray observations}
\label{Xobs}

We report here and in the following few sections on the multi-band data analysis and results of \rx, while in Section \ref{others} we report on the other X-ray sources in the field, in particular on the flaring X-ray source \tmsp.

\subsection{\xmm}
\label{xmm}

Two \xmm\ observations were carried out on 2015 October 20--21 (ID: 0764420101; PI: Rea) and December 
2--3 (ID: 0790180101; thanks to XMM Director's Time), respectively, and in both cases \rx\ was placed at the aim point of the European 
Photon Imaging Cameras (EPIC). The pn camera (Str\"{u}der et al. 2001) was configured 
in large window mode (LW; 47.7-ms time resolution), whereas the two MOS cameras 
(Turner et al. 2001) were operated in small window mode (SW; 0.3-s time resolution). 
The thin optical blocking filter was positioned in front of the cameras in both observations. 
The Reflection Grating Spectrometers (RGS; den Herder et al. 2001) were operated in standard 
spectroscopy mode. The journal of the two observations is reported in Table~\ref{tab:xray_log}.

\begin{table*}
\centering
\caption{Log of the X-ray observations.} 
\label{tab:xray_log}
\vspace{10pt}
\begin{threeparttable}
\begin{tabular}{ccccccccc} \hline
Satellite				& Obs. ID 			& Date 			& Start -- End time of exposure			& Instrument    		& Mode	 	& Exposure\tnote{a} & \rx\ count rate\tnote{b} \\
         				&				& 	(YY/MM/DD)	& (MJD)							&          	      		&		 	& (ks)			& (counts~s$^{-1}$)\\ 
\hline 
\multirow{5}{*}{\xmm}	& \multirow{5}{*}{0764420101}	&  \multirow{5}{*}{2015/10/20-21}  & \multirow{5}{*}{57\,315.4250 -- 57\,316.0116}		& EPIC pn			& LW	 	& 47.2 		& $2.053 \pm 0.007$\\
					&			         &  		&										& EPIC MOS\,1		& SW	 	& 51.1			& $0.5957 \pm0.0004$\\
					&				& 		&										& EPIC MOS\,2		& SW		& 51.0			& $0.5743 \pm 0.0003$\\
					&				& 		&										& RGS\,1			& spectroscopy	 & 52.7			& $0.0447 \pm 0.0007$\\	
					&				& 		&										& RGS\,2			& spectroscopy	 & 52.8			& $0.0456 \pm 0.0008$\\	
\hline
\multirow{5}{*}{\xmm}	& \multirow{5}{*}{0790180101}	& \multirow{5}{*}{2015/12/02-03} & \multirow{5}{*}{57\,358.9411 -- 57\,359.8111}		& EPIC pn 		& LW	 	& 69.1			& $1.714 \pm 0.007$\\
					&				& 		&										& EPIC MOS\,1		& SW	 	& 74.0			& $0.529 \pm 0.003$\\
					&				& 		&										& EPIC MOS\,2		& SW	 	& 74.0			& $0.533 \pm 0.003$\\	
					&				& 		&										& RGS\,1 			& spectroscopy	& 76.3			& $0.0422 \pm 0.0005$\\ 	
					&				& 		&									& RGS\,2 			& spectroscopy	& 76.4			& $0.0429 \pm 0.0007$\\ 	
\hline 
\multirow{14}{*}{\swift}	& 00041343002	& 2010/10/02 		&  55\,471.7469 -- 55\,471.9604				& XRT			& PC	 		& 4.1 			 & $0.086 \pm 0.005$\\
					& 00046948001	& 2011/12/18		&  55\,913.4285 -- 55\,913.5048				& XRT			& PC	 		& 2.0 			 & $0.074 \pm 0.006$\\
					& 00046948002	& 2014/10/07		&  56\,937.9180 -- 56\,937.9999				& XRT			& PC	 		& 3.0 			 & $0.069 \pm 0.005$\\
					& 00046948003	& 2014/10/11		&  56\,941.3212 -- 56\,941.8555				& XRT			& PC	 		& 2.6 			 & $0.061 \pm 0.005$\\
					& 00046948004	& 2014/10/15		&  56\,945.1198 -- 56\,945.5923				& XRT			& PC	 		& 2.7  			 & $0.082 \pm 0.006$\\
					& 00046948005	& 2014/10/19 		&  56\,949.2479 -- 56\,949.7930				& XRT			& PC	 		& 2.5 			 & $0.127 \pm 0.007$\\
					& 00041343003	& 2015/10/14		&  57\,309.4321 -- 57\,309.5680				& XRT			& PC 		& 2.0 			 & $0.18 \pm 0.01$\\
					& 00041343004	& 2015/12/27-28	&  57\,383.0380 -- 57\,384.1743				& XRT			& WT		& 10.4  			 & $0.217 \pm 0.005$\\
					& 00041343005	& 2015/12/31		&  57\,387.4330 -- 57\,387.6291				& XRT			& WT		& 1.4 			 & $0.37 \pm 0.02$\\				
					& 00046948006	& 2016/01/12		&  57\,399.0082 -- 57\,399.7305 				& XRT			& WT		& 4.0  			 & $0.312 \pm 0.009$\\
					& 00046948007	& 2016/01/13		&  57\,400.0037 -- 57\,400.5470				& XRT			& WT		& 0.5  			 & $0.13 \pm 0.02$\\
					& 00046948008	& 2016/01/14		& 57\,401.0678 -- 57\,402.0000					& XRT			& WT		& 3.5  			 & $0.151 \pm 0.007$\\
					& 00046948009	& 2016/01/17 		&  57\,404.3239 -- 57\,404.6413 				& XRT			& WT		& 2.8  			 & $0.157 \pm 0.008$\\
					& 00046948010	& 2016/01/20		& 57\,407.3096 -- 57\,407.8395					& XRT			& WT		& 3.7 			 & $0.190 \pm 0.008$\\
\hline						
\end{tabular}
\begin{tablenotes}
\item[a] Dead-time corrected on-source exposure.
\item[b] Average background-subtracted count rate in the 0.2--10~keV energy band for the EPIC instruments, in the 0.35--2.5~keV energy band for the combined RGS first order data sets, and in the 0.3--10~keV energy band for the \swift\ XRT data.
\end{tablenotes}
\end{threeparttable}
\end{table*}

\subsubsection{EPIC data}
\label{epic}

We processed the raw observation data files using the \textsc{epproc} (for pn data) and 
\textsc{emproc} (for MOS data) tasks of the \xmm\ Science Analysis System (\textsc{sas}\footnote{See \mbox{http://xmm.esac.esa.int/sas/}.}, 
version 15.0), with the calibration files in the database released in 2016 January (XMM-CCF-REL-332). 

The inspection of the EPIC light curves binned at 10~s revealed episodes of strong flaring particle background in both observations.
For the spatial and spectral analysis, we discarded the intervals affected by flaring background, which were pinpointed by applying intensity filters on the light curves, following the procedure described by De Luca \& Molendi (2004). This reduced the effective exposure time to approximately 38.9 (pn), 48.5 
(MOS\,1) and 48.4 (MOS\,2) ks for the first observation, and to 64.6 (pn), 71.5 (MOS\,1) and 71.2 (MOS\,2) ks for the second observation.

\begin{figure}
\centering
\includegraphics[width=6cm,angle=-90]{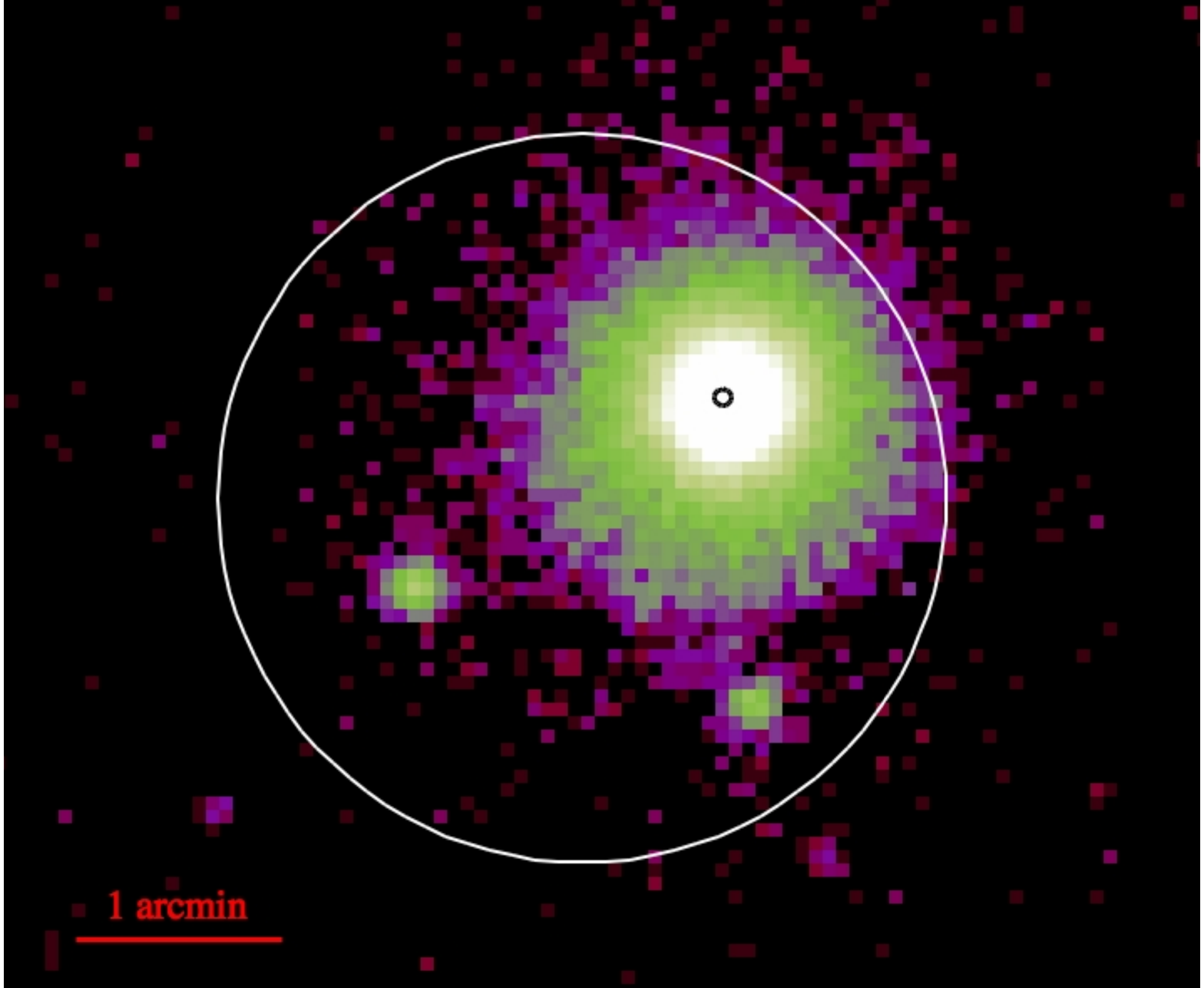}
\caption{Field around \rx\, in the 0.2--10\,keV band as observed by \xmm. The image shows the merging of all observations and EPIC instruments, for a total exposure time of $\sim$125\,ks. The white circle shows the position of the Fermi-LAT source \fermirx\, with a 0.03\deg\ error radius (at 68\% confidence level) as reported in the 3FGL catalog. The black circle is centred on the best optical position, with an error circle of 2 arcsec (enlarging by a factor of 10 the optical positional accuracy for imaging purposes). North is up, and east is left.}
\label{xmmfov}
\end{figure}

We ran the \textsc{sas} source detection script \textsc{edetect$\_$chain} on the combined event 
lists from the two observations. 
This returned for \rx\ the position $\rm RA=08^h38^m43\fs20$, $\rm Dec= -28^\circ27'01\farcs44$ (J2000.0), with an  
uncertainty of 1.4 arcsec (radius, at the 90 per cent confidence level).\footnote{The uncertainty 
on the position is evaluated as the quadratic sum of the statistical uncertainty given by the 
task \textsc{edetect$_-$chain} and the systematic absolute pointing uncertainty of \xmm.} 
This is fully consistent with the more accurate position of the optical counterpart, 
$\rm RA=08^h38^m43\fs37$, $\rm Dec= -28^\circ27'01\farcs5$, with an uncertainty of about 
0.2~arcsec (Masetti et al. 2013). 
The 0.2--10~keV image of the field of \fermirx\, is shown in Fig.~\ref{xmmfov}, and \rx's light-curves in Fig.\,\ref{xmmlc}. It was created by stacking the images of the EPIC cameras cleaned from bad pixels and hot columns from both observations. Apart from \rx, two other X-ray sources were identified in the 68 per cent error circle of \fermirx . In the following we will dub them as \tmsp\ and \thirdxmm\ (see Table~\ref{tab:log_sources} for details on their position and their X-ray properties).

For the study of \rx, we collected the source counts within a circle with a 40 arcsec radius for the pn data, and 30 arcsec for the MOS data. The background was extracted from circles located on the same CCD as the source (so as to guarantee similar low-energy noise subtraction) and avoiding detector areas contaminated by possible out-of-time events from the source (in the pn) or too near to the CCD edges. On the other hand, data products for the other two sources were extracted only from the event lists of the pn, because both fell outside the central window of the MOS instruments.

We restricted the analysis to photons in the 0.2--10~keV energy range with \textsc{pattern} $\leq4$ for the pn and $\leq12$ for the MOS, and \textsc{flag} = 0 (we also checked that the data were not affected by pile-up).
For the spectral analysis of \rx\ we used only the data acquired by the pn camera, which provides the spectra with the highest counting statistics, to avoid systematic errors introduced by cross-calibration uncertainties (we checked that the MOS spectra are in agreement with the results from the pn camera). We generated the spectral redistribution matrices and ancillary response files for each of the three sources with the \textsc{rmfgen} and \textsc{arfgen} tools, respectively.  
The fits were performed with the \textsc{xspec} package (version 12.9.0; Arnaud 1996), using the $\chi^2$ statistics. For the timing analysis, we converted the photon arrival times to Solar system barycentre reference frame using the \textsc{sas} task \textsc{barycen} and the DE-200 Solar System ephemeris.

\subsubsection{RGS data}
\label{rgs}

\begin{figure}
\begin{center}
\includegraphics[width=8cm]{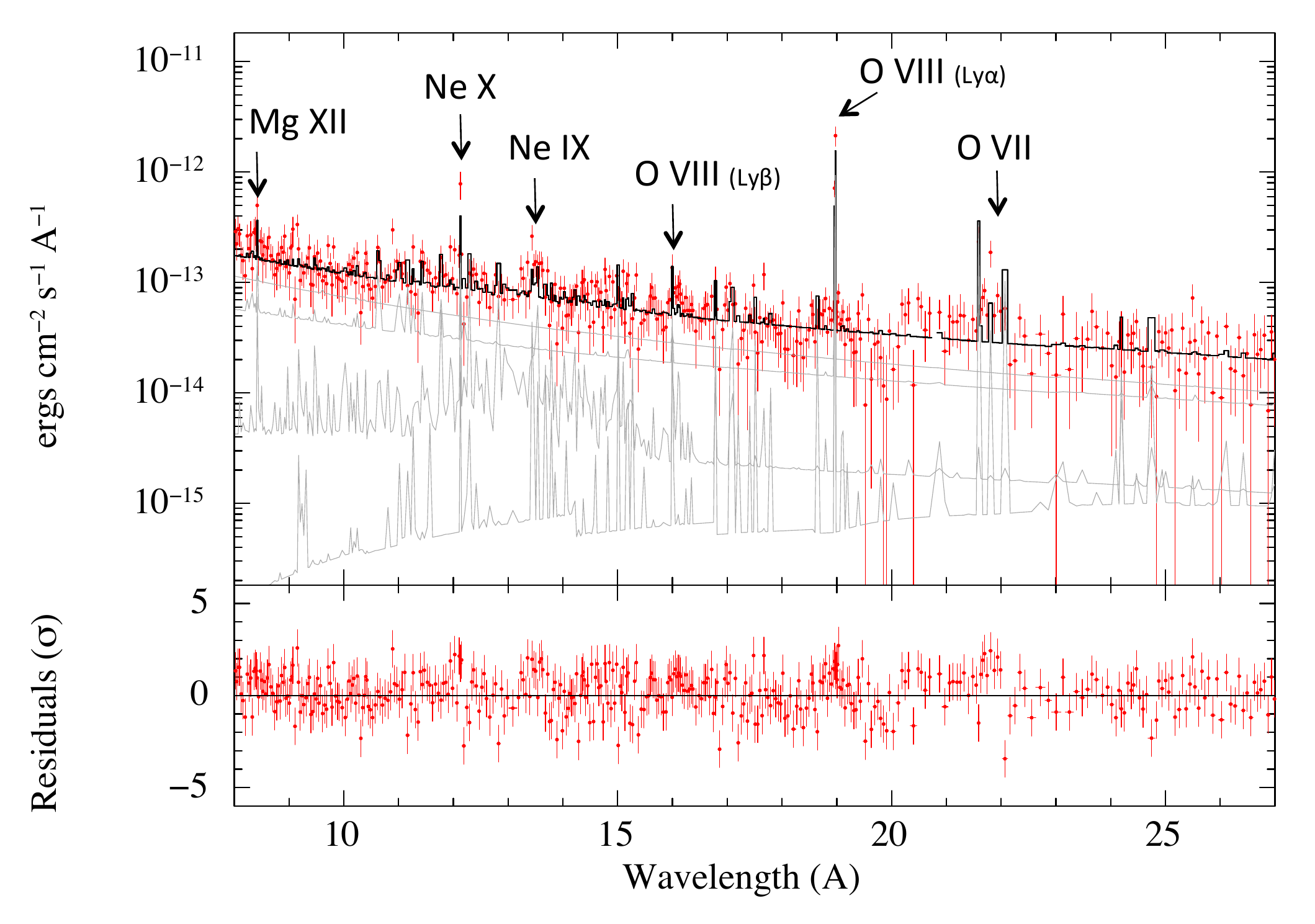}
\caption{\xmm\ RGS first order spectra of \rx\, for the second observation. Grey lines represent the three ionized plasma components of the model (black line) that best fits the pn (and RGS) data.}
\label{fig:rgs}
\end{center}
\end{figure}

We processed the data using the task \textsc{rgsproc} of \textsc{sas} to produce calibrated and 
concatenated photon event lists, spectra and response matrices. We focused our analysis on the 
first order spectra alone, which are the best calibrated and have a higher number of counts. First, 
we subtracted the model background spectra, which were created by the standard RGS pipeline 
and are template background files based on the count rate in CCD 9. We then combined the spectra 
using \textsc{rgscombine}, which appropriately accounts for the response matrices and backgrounds 
of the different spectrometers, and rebinned the resulting spectra to have at least one count per 
energy bin. The analysis was limited to the 0.35--2.5~keV energy interval (5--38 \AA), where 
the calibration of the response is most accurate (see Fig.\,\ref{fig:rgs}).
.


\subsection{\swift\ X-Ray Telescope}
\label{swift}

\begin{figure}
\begin{center}
\includegraphics[width=8cm,height=5cm]{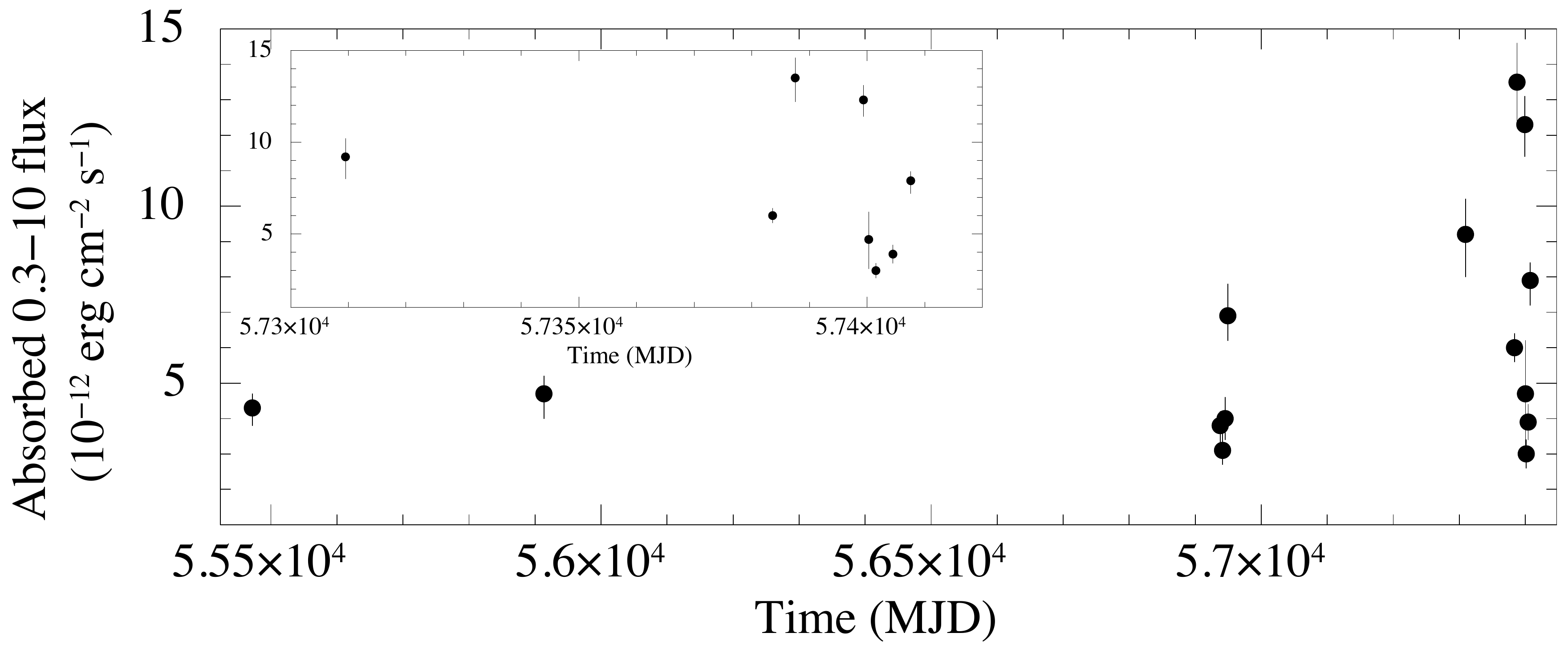}
\caption{\swift\ XRT long-term light curve of \rx. Fluxes relative to the most recent observations 
are shown in the inset.}
\label{fig:xrt_lcurve}
\end{center}
\end{figure}

The X-ray Telescope (XRT; Burrows et al. 2005) on board the \swift\ satellite observed the field 
of \rx\ for a total of 14 times between 2010 September 8 and 2016 January 20 (see Table~\ref{tab:xray_log} 
for a log of the observations, and Fig.\,\ref{fig:xrt_lcurve}). The first 7 observations were carried out in photon counting 
(PC) mode, which provides a two-dimensional image and a time resolution of 2.5073~s. The last 
7 observations were instead performed in windowed timing (WT) mode, i.e., data were transferred 
from the imaging array to the frame store array with an integration time of 1.78~ms, at the 
expense of one dimension of spatial resolution (see Hill et al. 2004 for a detailed description of 
the XRT readout modes). 

The data were analysed primarily to study the X-ray long-term variability of \rx. 
We processed the data with standard screening criteria 
and generated exposure maps with the task \textsc{xrtpipeline} (version 0.13.2) from the \textsc{ftools} 
package (Blackburn 1995), using the optical position of the source (Masetti et al. 2013) and the 
spacecraft attitude file. We selected events with grades 0--12 and 0 for the PC and WT data, 
respectively, and extracted the source and background spectra using \textsc{xselect} (v. 2.4). We 
accumulated the source counts from a circular region with a radius of 20 pixels (one XRT pixel 
corresponds to about 2.36 arcsec). To estimate the background in the PC-mode data, we extracted 
the events within an annulus centred on the source position with inner and outer radius of 40 and 
80 pixels, respectively. For the WT-mode data we opted for a circle far from the target and of the 
same size as the source one, owing to the small extent of the WT window.

We created the observation-specific ancillary response files (using exposure maps) with \textsc{xrtmkarf} 
(v. 0.6.3), which corrects for the loss of counts due to hot columns and bad pixels, and accounts for 
different extraction regions, telescope vignetting and PSF corrections. We then assigned the appropriate 
redistribution matrix available in the \textsc{heasarc} calibration data base,\footnote{See 
http://www.swift.ac.uk/analysis/xrt/rmfarf.php.} excluded bad spectral channels, and grouped 
the background-subtracted spectra to have at least 20 counts in each spectral bin. We limited 
the spectral analysis to the 0.3--10~keV energy range for PC-mode data and to the 0.7--10~keV interval 
for the WT-mode data, owing to known calibration issues in WT that result in spurious bumps 
and/or turn-ups in the spectra at low energy.\footnote{See http://www.swift.ac.uk/analysis/xrt/digest$_-$cal.php.}

\subsection{INTEGRAL}
\label{integral}

We analyzed all publicly available {\it INTEGRAL} data collected 
in the direction of the source in the past 13~yrs of the mission. We considered all {\it INTEGRAL} science windows (ScW), 
i.e. the different pointings lasting each $\sim$2--3\,ks, in which the source was 
observed within 12~degrees off axis from the satellite aim point for the IBIS/ISGRI 
instrument (Ubertini et al. 2003; Lebrun et al. 2003) and 3.5~degrees off axis for the two JEM-X instruments (Lund et al. 2003). 
All data were processed and analyzed by making use of the Off-line Scientific Analysis software 
(OSA) version 10.2 distributed by the ISDC (Courvoisier et al. 2003).

We found that the region around \fermirx\, was probed by {\it INTEGRAL} for a total of 892~ks 
effective exposure with IBIS/ISGRI, and 15.5~ks (10.0~ks) with JEM-X1 (JEM-X2). 
The dataset spanned from the satellite revolution 28 (starting on 52644~MJD) to 1625 (ending on 57381~MJD). 
We built the IBIS/ISGRI mosaics in the two 20--40~keV and 40--80~keV energy bands, as well as the JEM-X 
mosaics in the 3-10~keV energy band. \rx\, was not detected by these instruments. We estimated with the 
OSA {\tt mosaic\_spec} tool a 3$\sigma$ upper limit on the source hard X-ray flux of 1~mCrab  
(roughly 8$\times$10$^{-12}$~erg~cm$^{-2}$~s$^{-1}$) in the 20--40 keV energy band and 1.5~mCrab 
(roughly 10$^{-11}$~erg~cm$^{-2}$~s$^{-1}$) in the 40--80 keV energy band. The corresponding 
upper limit in the 3--10~keV energy band was of 5~mCrab (roughly 7$\times$10$^{-11}$~erg~cm$^{-2}$~s$^{-1}$), consistent with the \xmm\ findings.

\begin{table}
\caption{Results of the fits of the \xmm\ EPIC pn spectra of \rx\, at the minimum and maximum of the $\sim$54.7\,ks beat modulation. For each beat phase the spectra of both observations 
were fitted together to the \textsc{tbabs*pcfabs*(mekal+mekal+mekal)} model, with all normalizations free to vary. Uncertainties are quoted at the 90\% confidence level for a single parameter of interest.}
\label{tab:spectral_parameters}
\centering
\begin{tabular}{@{}lcc}
\hline
Parameter										& Minimum & Maximum  \\
\hline \vspace {0.1cm}
$\nh^a$  			($10^{20}$ cm$^{-2}$) 	  & 		 $2.3_{-0.6}^{+0.8}$      	& $2.4\pm0.5$ 		\\ 	\vspace {0.1cm}

Pcfabs $\nh$   ($10^{24}$ cm$^{-2}$) 	   		& $1.8_{-0.7}^{+1.3}$     	&  $2.9_{-1.1}^{+1.8}$		\\ 	\vspace {0.1cm}

Covering fraction (\%) & $72_{-24}^{+18}$ &  $67_{-30}^{+23}$       \\ 	\vspace {0.1cm}

A$_Z$ 			(A$_{Z,\odot}$)					& $0.2_{-0.1}^{+0.2}$	& $0.37 \pm 0.1$			\\ 	\vspace {0.1cm}

$kT_1$			(keV)						& $0.22\pm0.03$	& $0.12\pm0.02$			\\ 	\vspace {0.1cm}	
Normalization$^b$					& $7.1_{-0.3}^{+1.8}$ (obs 1)	& $30.2_{-1.8}^{+6.7}$	(obs 1)	\\ 	\vspace {0.1cm}
											& $5.2_{-0.3}^{+1.4}$ (obs 2)	& $21.3_{-1.3}^{+4.7}$	(obs 2)	\\ 	\vspace {0.1cm}
$kT_2$			(keV)						& $1.03\pm0.14$		& $1.00\pm0.08$		\\ 	\vspace {0.1cm}	
Normalization					& $8.9_{-0.6}^{+2.4}$ (obs 1)	& $4.2_{-0.4}^{+1.7}$ (obs 1)		\\ 	\vspace {0.1cm}
											& $6.5_{-0.5}^{+1.9}$ (obs 2)		& $7.6_{-0.3}^{+0.6}$  (obs 2)	\\ 	\vspace {0.1cm}
$kT_3$			(keV)						& $11.7_{-2.6}^{+7.3}$	& $14.8_{-0.9}^{+1.2}$		\\ 	\vspace {0.1cm}	
Normalization					& $38_{-1.7}^{+7.7}$ (obs 1)	& $161_{-8}^{+31}$ 	(obs 1)			\\ 	\vspace {0.1cm}
											& $34_{-1.6}^{+6.9}$	  (obs 2)    & $130_{-6}^{+13}$  (obs 2)				\\ 	\vspace {0.1cm}
\multirow{2}{*}{Absorbed flux$^c$ }		& $2.3\pm 0.7$ (obs 1)     & $10.3 \pm 1.0$ (obs 1)   	 	\\	\vspace {0.1cm}
											& $2.0\pm0.7$ (obs 2)    	 &	$8.4 \pm 1.2$ (obs 2) \\ \vspace {0.1cm}
$\chi^2_{\nu}$ 	     (dof)        				         	& 1.09 (213) 		& 1.06 (541)      	     		\\	\vspace {0.1cm}
Null hypothesis probability							& $1.5 \times 10^{-1}$    &  $1.3 \times 10^{-1}$	    		 \\	
\hline
\end{tabular}  
\begin{list}{}{}
\item[$^{a}$] The abundances are those of Wilms et al. (2000). The photoelectric absorption cross-sections are from Verner et al. (1996).
\item[$^{b}$] Normalizations are in units of $10^{-4}$ cm$^{-5}$.	
\item[$^{e}$] Fluxes are in units of $10^{-12}$ \flux and in the 0.2--10~keV energy range. 
\end{list}
\end{table}

\section{Infrared and Optical observations} 
\label{opt}

\subsection{OAGH}

Near-IR observations of the field of the \fermirx\, source were carried out on 2016 January 14 with the 2.1-m telescope 
of the Observatorio Astrofisico Guillermo Haro (OAGH) in Cananea, Mexico. The telescope is equipped 
with the CAnanea Near Infrared CAmera (CANICA), which provides an image scale of 0.32 arcsec/pixel, 
and covers a field of view of about $4 \times 4$ arcmin$^2$. CANICA observed with the $H$ filter starting 
at 08:03:46 UT for an exposure of 60~s, and with the $K$ filter starting at 08:13:56 UT for an exposure of 
30~s. The counterparts to the three X-ray sources were detected in both bands, with magnitudes: $H=17.16 \pm 0.2$ and $K=16.18 \pm 0.09$ for \rx, $H= 17.0\pm0.2$ and $K=16.7\pm0.1$ for \tmsp, and $H=17.0\pm0.2$ and $K=16.2\pm0.1$ for \thirdxmm . The zero points were 20.18 and 21.34 for the $H$ and $K$ filters, respectively (see also Fig.\,\ref{iacfov}). Neither \tmsp\ nor \thirdxmm\ have a counterpart in the Two-Micron All-Sky Survey (2MASS) catalogue\footnote{See http://www.ipac.caltech.edu/2mass/releases/second/.}.

\subsection{XMM-Newton Optical Monitor}
\label{om}

The \xmm\ Optical/UV Monitor Telescope (OM; Mason et al. 2001) was configured in `Image' mode throughout 
the first observation and used the $V$ filter, providing a wavelength coverage within the 5100--5800~\AA\ range.
A total of 10 exposures were acquired, resulting in a total dead-time-corrected on source exposure time of about 
45.8~ks (the first 5 exposures had a duration of about 5~ks each, whereas all the following 5 exposures lasted 
about 4.16~ks). The OM acquired the data in both `fast-window' and `Image' modes during the second 
observation, using the $B$ filter (centred on 4392 \AA\ and with band pass between 3800 and 5000 \AA). No useful 
analysis could be carried out on the data sets taken in fast-window mode, because the source fell outside the small 
square window (23$\times$23\,px$^2$) of the instrument. We then focused on the data in the `Image' mode, which 
consist of 15 4.4-ks frames for a total exposure of 66.0~ks.

\begin{figure}
\begin{center}
\includegraphics[width=8.5cm]{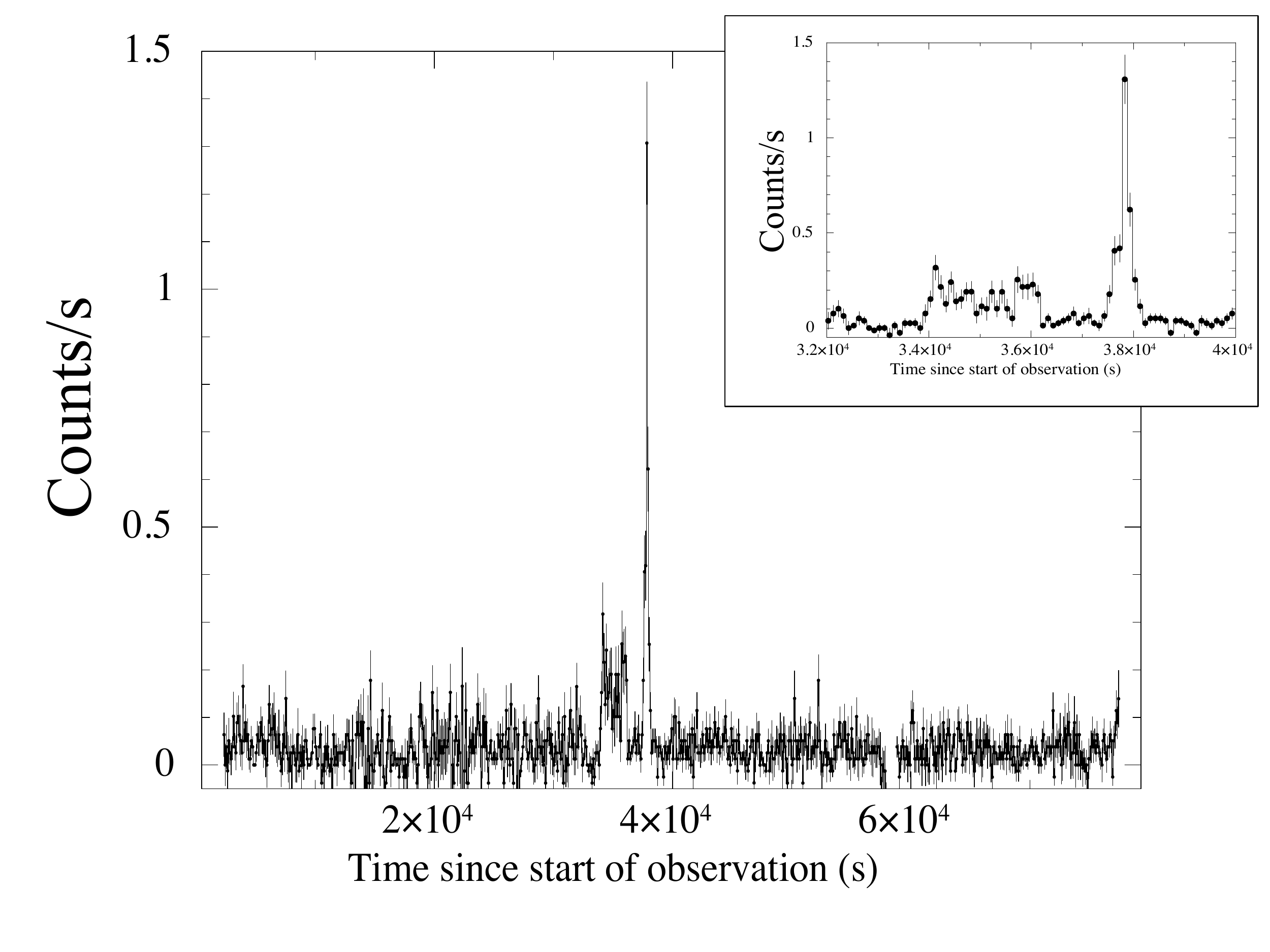}
\includegraphics[width=8.5cm]{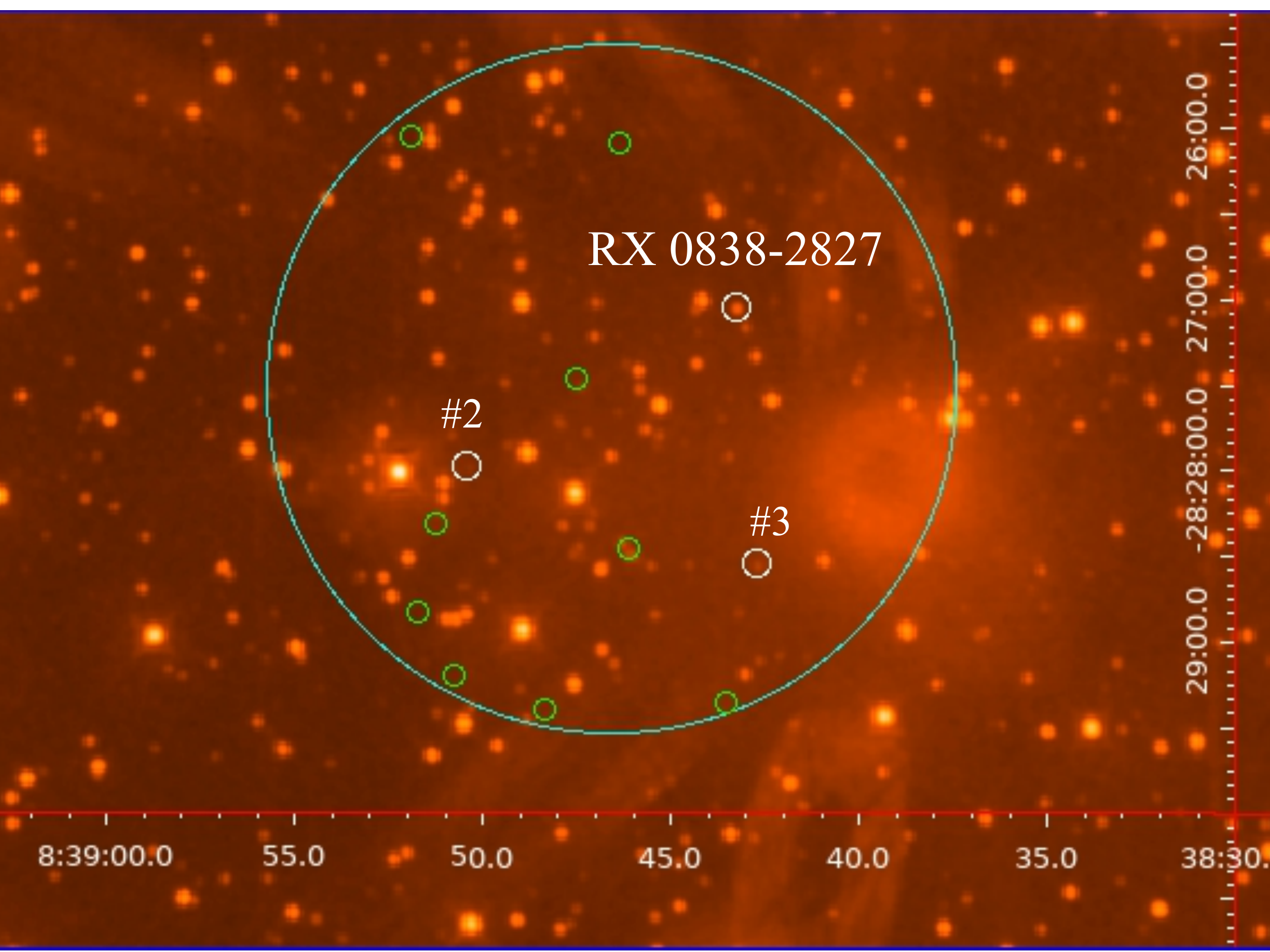}
\caption{{\em Top panel}: 0.2--10~keV background-subtracted and exposure-corrected pn light curve of source \#2 (\tmsp) during 
the second \xmm\ observation. A binning time of 400~s was adopted in the inset with the zoom of the flare. Time is 
in units of seconds since the start of the exposure. {\em Bottom panel}: XMM-OM field of view with superimposed the 68\% {\it Fermi}-LAT position of 3FGL\,J0838.8$-$2829 (light blue), the three X-ray sources detected in the XMM-pn field of view (white), and in green we report on the position of all the radio detection at 5.5\,GHz within the 3FGL error circle (see text for details). }
\label{src2_flare}
\end{center}
\end{figure}

We extracted the background-subtracted photometric data using the \textsc{omichain} processing pipeline with 
the default parameter settings, as recommended by the \textsc{sas} threads\footnote{See 
http://xmm2.esac.esa.int/sas/8.0.0/documentation/threads/omi$_-$thread.html}. Only one source is detected within 
the error circle of the X-ray position of \rx, for which we show the light curve in Fig.\,\ref{xmmlc}. The optical emission 
clearly shows a variability correlated with the X-ray emission, around an average value of 17.77 mag in the $V$ 
filter and 17.53 mag in the $B$ filter (expressed in the Vega photometric system). \tmsp\ and \thirdxmm\
are clearly detected in the second, longest observation. Interestingly, the former exhibits a flare ($B \sim$ 19.5 
mag) simultaneous to that observed in the soft X-ray flux (see Section\,\ref{others} and 
Fig.\,\ref{src2_flare}). No significant flaring episodes are instead observed in the 
data sets of the latter source.

\begin{table}
\caption{Log of IAC80 imaging observations and results of photometry of the source \tmsp . Magnitudes are in the Vega system and are not corrected for Galactic extinction.}
\begin{tabular}{cccc} \hline 
  mid exp time       & texp             &Filter &   mag  	        \\
    (MJD)            & (s)              &       &		        \\ \hline
\multicolumn{4}{|c|}{$XMM J083850.4-282759$}                                                                                 \\ \hline
 57367.21949	     & $18 \times 300$s & $B$	&   $22.20 \pm 0.17$	\\
 57367.22329	     & $18 \times 300$s & $R$	&   $19.83 \pm 0.12$	\\ 
 57368.20316	     & $16 \times 300$s & $B$	&   $21.86 \pm 0.13$	\\
 57368.20314	     & $16 \times 300$s & $R$	&   $19.71 \pm 0.12$	\\ 
 57392.11902	     & $11 \times 300$s & $B$	&   $21.31 \pm 0.11$	\\	     
 57392.12283	     & $12 \times 300$s & $R$	&   $19.55 \pm 0.12$	\\		     
 57415.07043	     & $15 \times 300$s & $B$	&   $>21.5$ ($3\sigma$ U.L.)	\\		   
 57415.06659	     & $16 \times 300$s & $R$	&   $20.08 \pm 0.15$	\\ \hline		  \hline
\hline
\end{tabular}   
\label{tab_log_iac80_MSP}
\end{table}

\subsection{IAC 80}
Optical imaging observations of the field of 3FGLJ0838.8-2829 were carried out with the IAC-80 telescope on 2015 December 11 and 12 (MJD 57367 and 57368), on 2016 January 5 and 28 (MJD 57392 and 57415). A set of exposures lasting 300 $s$ were obtained on each night alternating the $B$ and $R$ filters. Image reduction was carried out by following the standard procedures: subtraction of an averaged bias frame, division by a normalized flat frame. Astrometry was performed using the USNOB1.0\footnote{http://www.nofs.navy.mil/data/fchpix/} catalogue. Aperture photometry was made with the PHOTOM software part of the STARLINK\footnote{http://starlink.eao.hawaii.edu/starlink} package. The photometric calibration was done against the APASS catalogue. In order to minimise any systematic effect, we performed differential photometry with respect to a selection of local isolated and non-saturated standard stars. In Fig.\,\ref{iacfov} we show two images taken on December 11 in the $B$ and $R$ bands.
We detected the three X-ray sources in both bands. \rx\, displays significant variability matching the periodicity observed in the X-rays and with the OM. The mean magnitudes are $B \sim 17.7$ and $R \sim 17.0$. The other two X-ray sources are detected with mean magnitudes $B \sim 21.7$ and $R \sim 19.8$ for \tmsp\ and $B \sim 20.0$ and $R \sim 19.01$ for \thirdxmm . This last source shows a steady optical flux. On the other hand, some variability is revealed in \tmsp\, both in the single $R-$band exposures (in the $B-$band the source is too faint to be detected in the single frames) and among the different nights (Tab.~\ref{tab_log_iac80_MSP}).

\subsection{ESO NTT observations}

\begin{figure}
\centering
\vbox{
\includegraphics[width=8cm,angle=0]{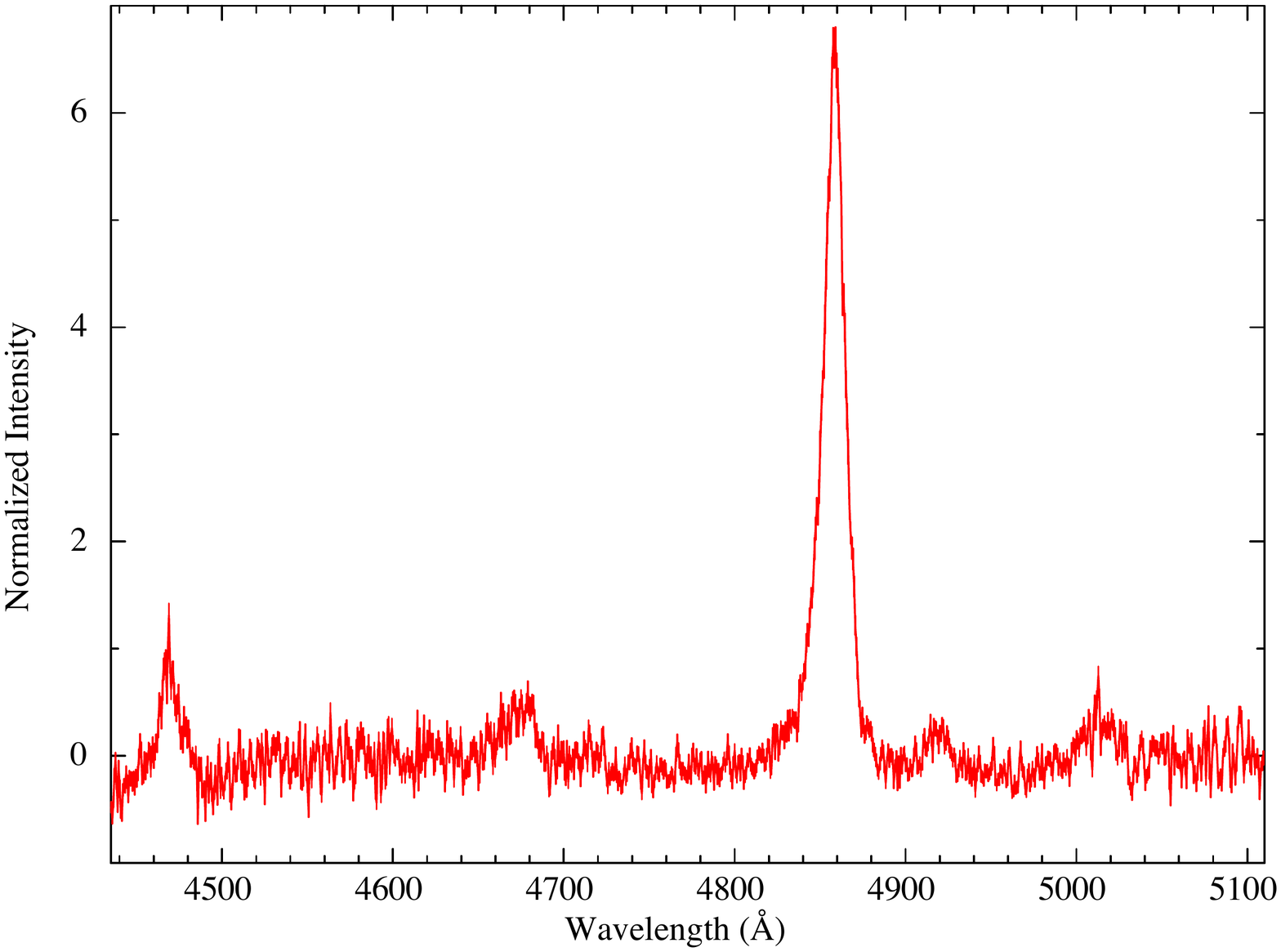}
\includegraphics[width=8cm,angle=0]{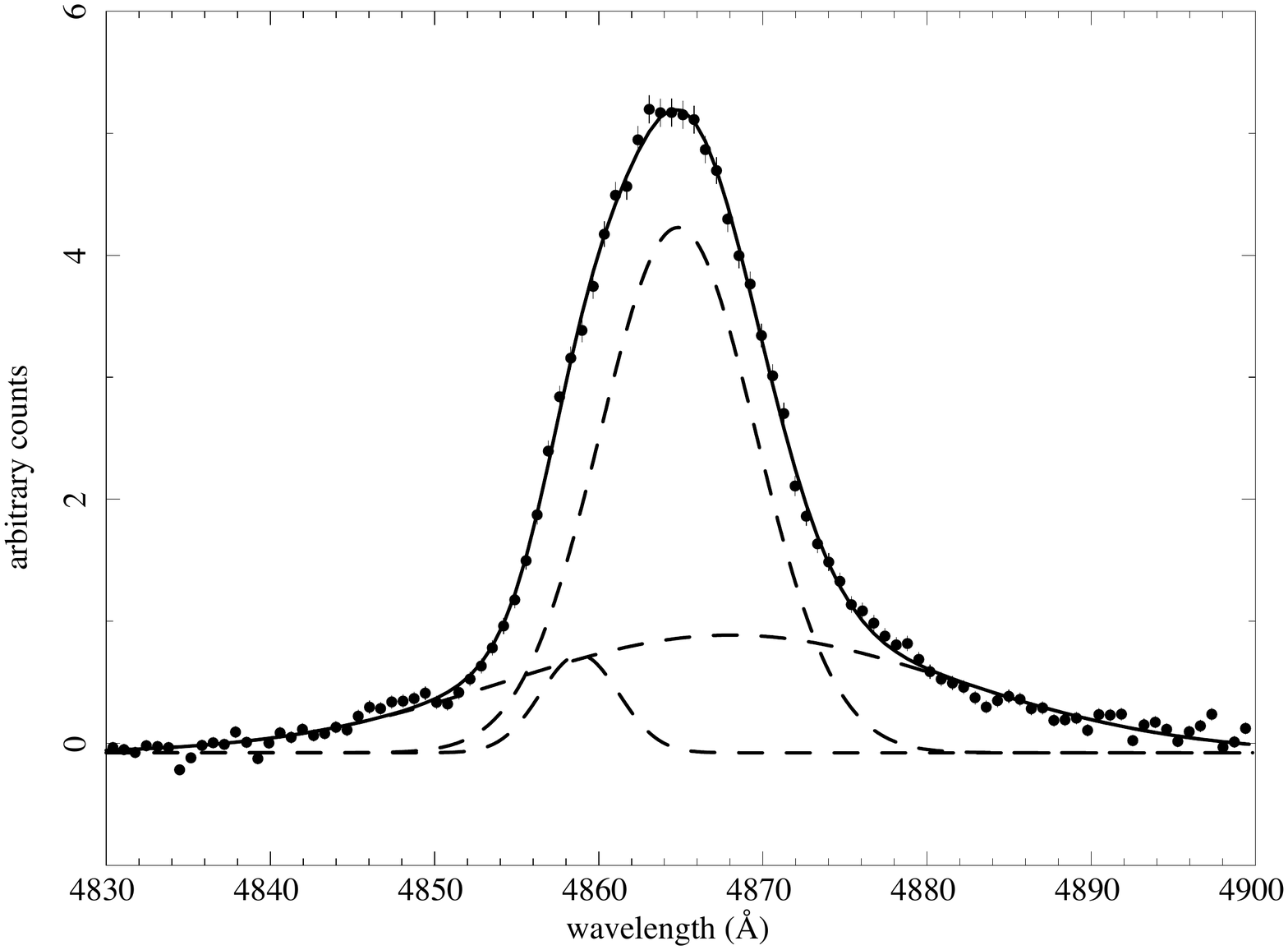}}
\caption{\label{nttspectrum} Top panel: an example of the ESO-NTT spectra of \rx. Bottom panel: modelling of the H$\beta$ emission line with three Gaussians (see text for details).}
\end{figure}

We performed optical spectral observations of \rx\, on 2016 February 27 with the ESO-NTT telescope (ID: 296.D-5034; PI: Rea). The observations were carried out with the EFOSC2 spectrograph using grism 19 and a slit aperture of $1''$,  covering the spectral range 4445--5110 \AA\, with resolution $R \sim 2500$ (0.67 \AA~pixel$^{-1}$). A total of 29 spectra, each one lasting 550 s, were obtained in the time interval from 01:05:30 UT--06:55:33 UT. Data were reduced using standard procedures using the ESO-MIDAS package for bias subtraction and flat-field correction. Wavelength calibration was carried out using helium-argon lamps. We show in Fig.\,\ref{nttspectrum} and \ref{ntt} the acquired spectra.

\section{Radio observations} 
\label{radio}

\begin{figure*}
\centering
\vbox{
\includegraphics[width=12cm,angle=0]{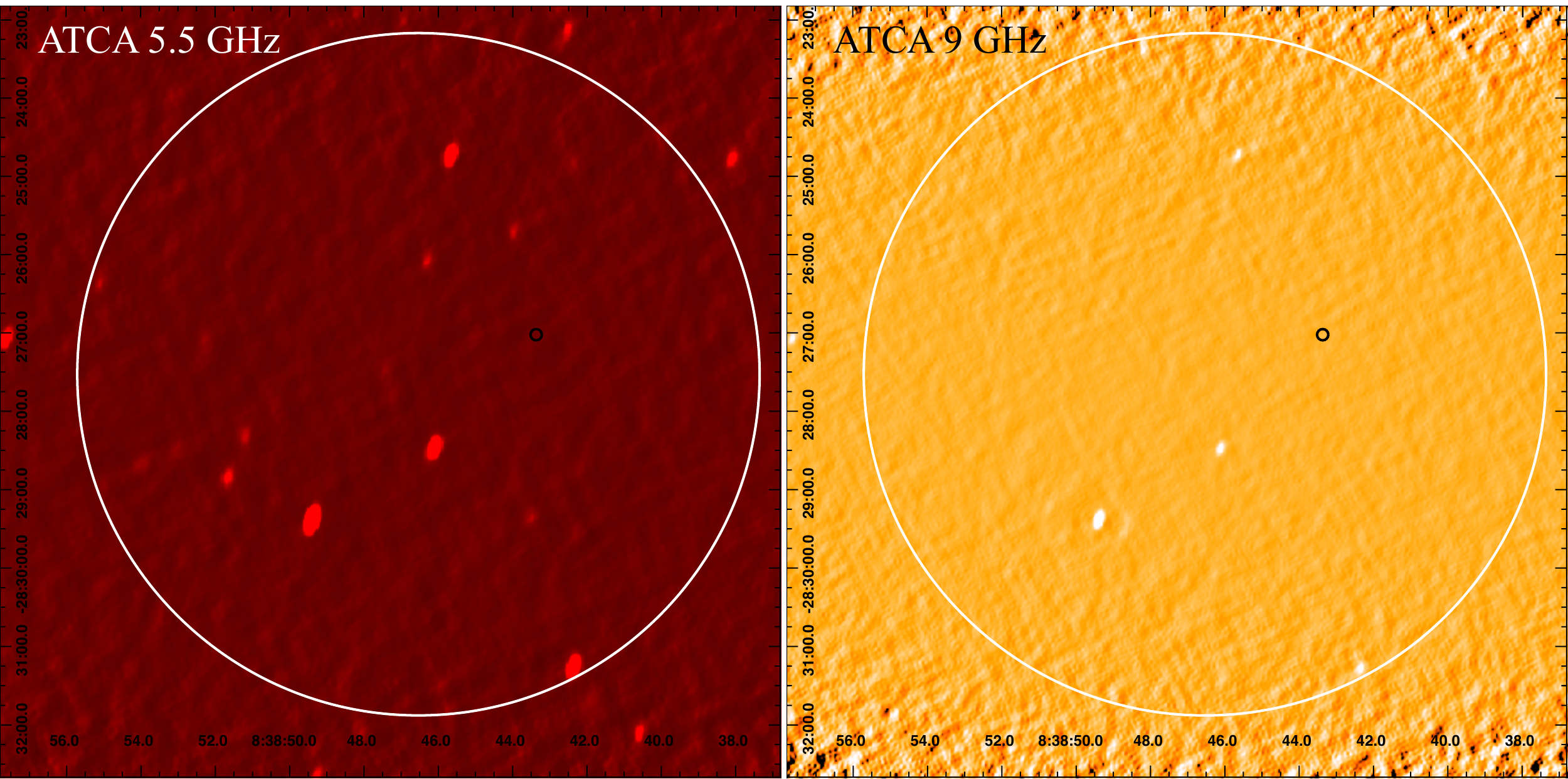}
\includegraphics[width=12cm,angle=0]{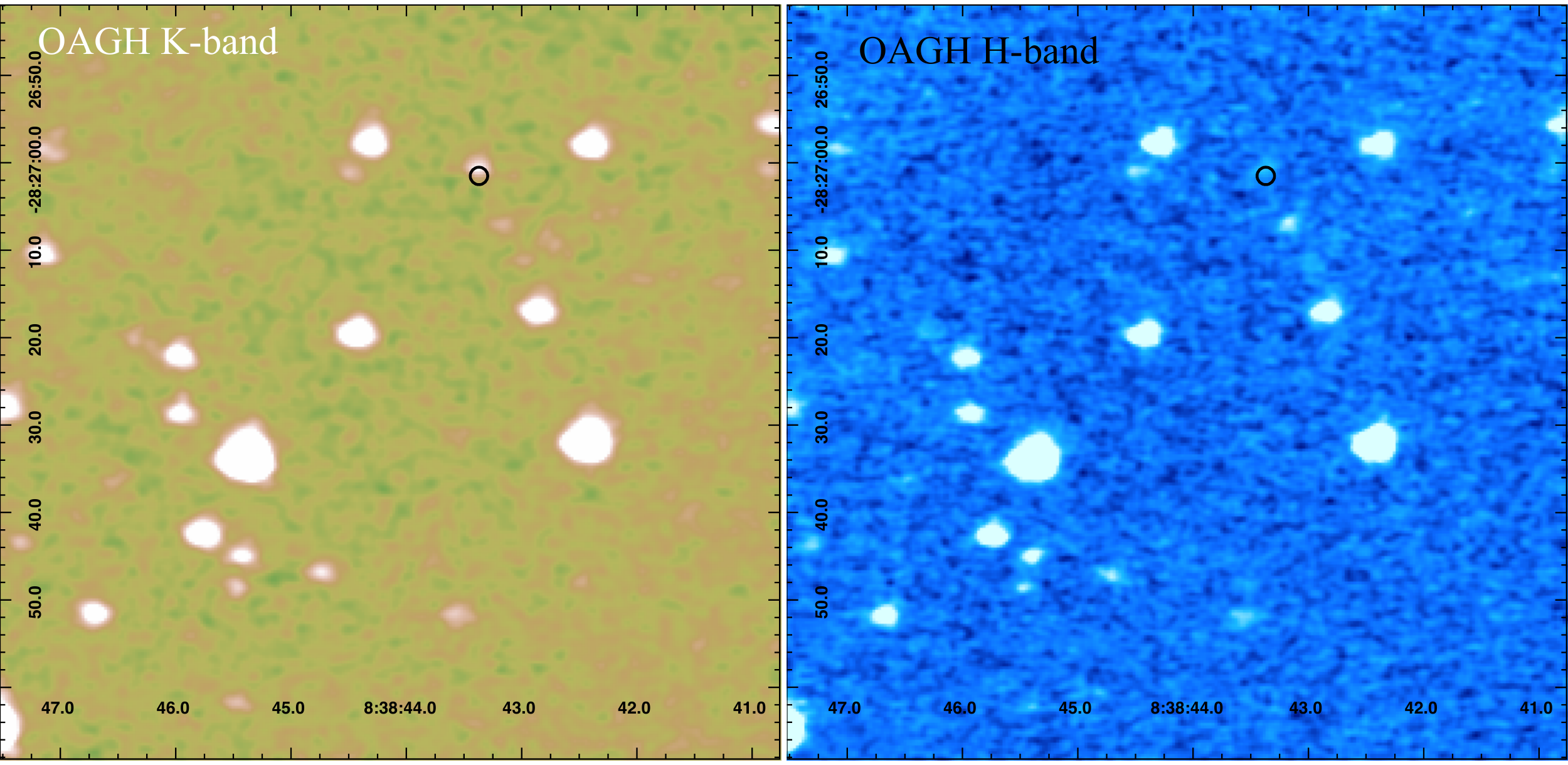}
\includegraphics[width=12cm,angle=0]{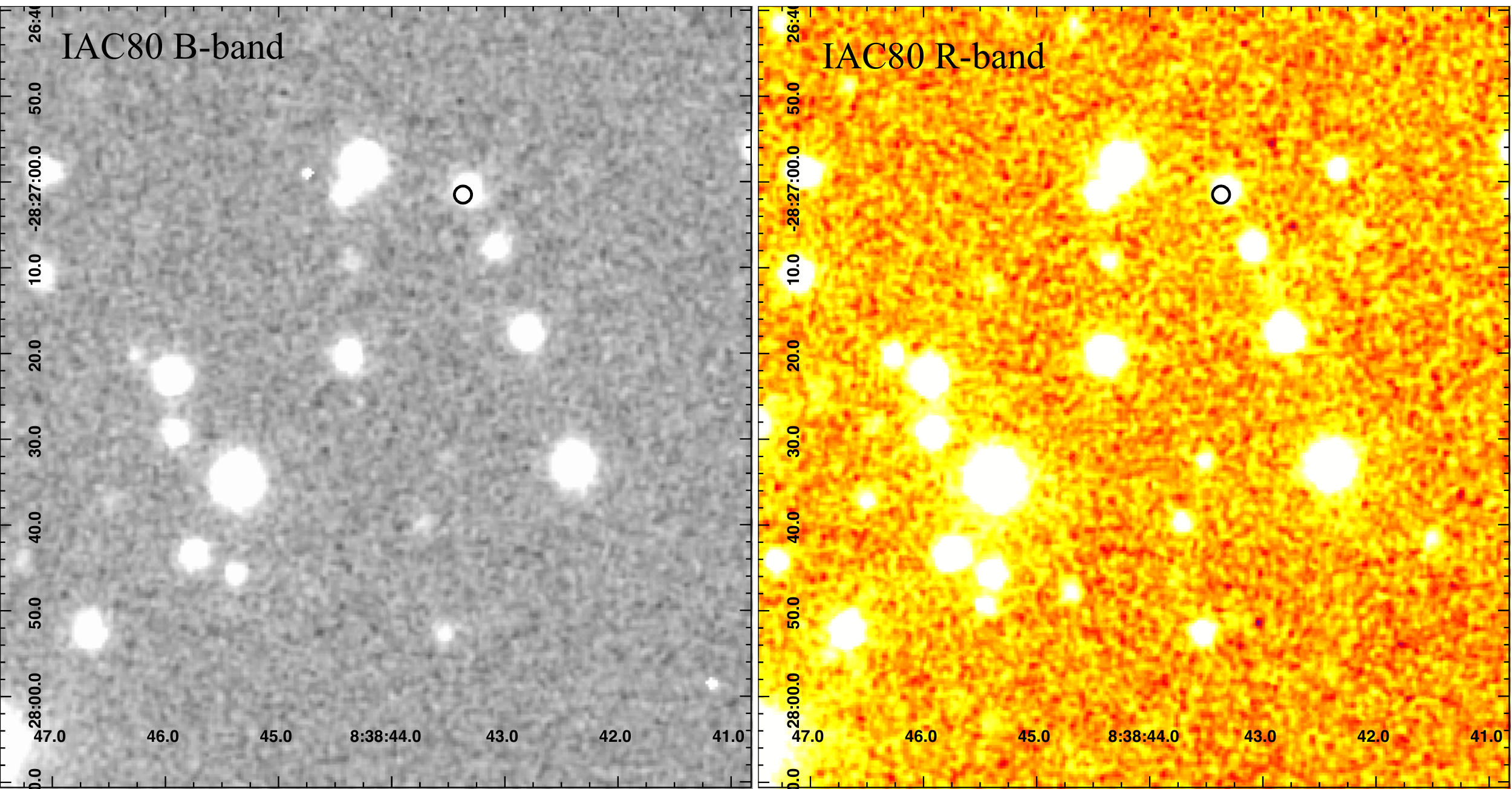}}
\caption{Field around \rx\, in different energy bands. From top-left to bottom-right we show the radio images taken by ATCA at 5.5 GHz (top left) and 9 GHz (top right) with superimposed the Fermi-LAT 3FGL position (0.033deg error circle at 68\% confidence level), and a 2 arcsecond black circle around the optical position of \rx. In the middle panels we show the infrared $K$ band (middle left) and $H$ band (middle right) as observed by the OAGH telescope, and in the bottom panels the optical $B$ band (bottom left) and $R$ band (bottom right) from the IAC80 telescope. The black circle of the middle and bottom panels is centred on the \rx\, best optical position with an error radius of 1 arcsec (enlarging by a factor of 5 the optical positional accuracy for imaging purposes). North is up, and east is left.}\label{iacfov}
\end{figure*}

We observed the field around \fermirx's position using the \emph{Australia Telescope Compact Array} in the 1.5A configuration at 5.5 and 9~GHz with 
a bandwidth of 2~GHz at each frequency (TOO project CX339). The observations were carried out on
2015 December 4 from 12:46 to 20:28 (UT), with a total integration time of 6.2 hours. We used 
PKS\,B1934-638 as the flux calibrator and 0858-279 as the phase calibrator. After standard calibration 
and imaging with \textsc{miriad} (Sault, Teuben \& Wright 1995), we ran one round of phase self-calibration 
and one round of complex amplitude self-calibration. The resulting images (with robust =0.5 weighting) 
have an rms of 6.3 and 7.1~$\mu$Jy at 5.5 and 9~GHz (see Fig.\,\ref{iacfov}). The synthesised beam sizes are 11.1 arcsec by 2.3
arcsec (P.A. --6.5\deg) and 7.35 arcsec by 1.75 arcsec (P.A. --6.5\deg), respectively. We detect no radio 
source at the position of \rx\, or at those of the other two X-ray sources down to a 3$\sigma$ upper limit of $\sim$$20$~$\mu$Jy at both frequencies. Other unidentified sources are detected in the field (see Fig.\ref{iacfov} and \ref{src2_flare}).

\section{\rx: data analysis and results}
\label{analysis}

\subsection{X-ray timing analysis} 
\label{timing}
\label{xmmtiming}

\begin{figure*}
\centering
\resizebox{\hsize}{!}{\includegraphics[width=14cm,height=8cm]{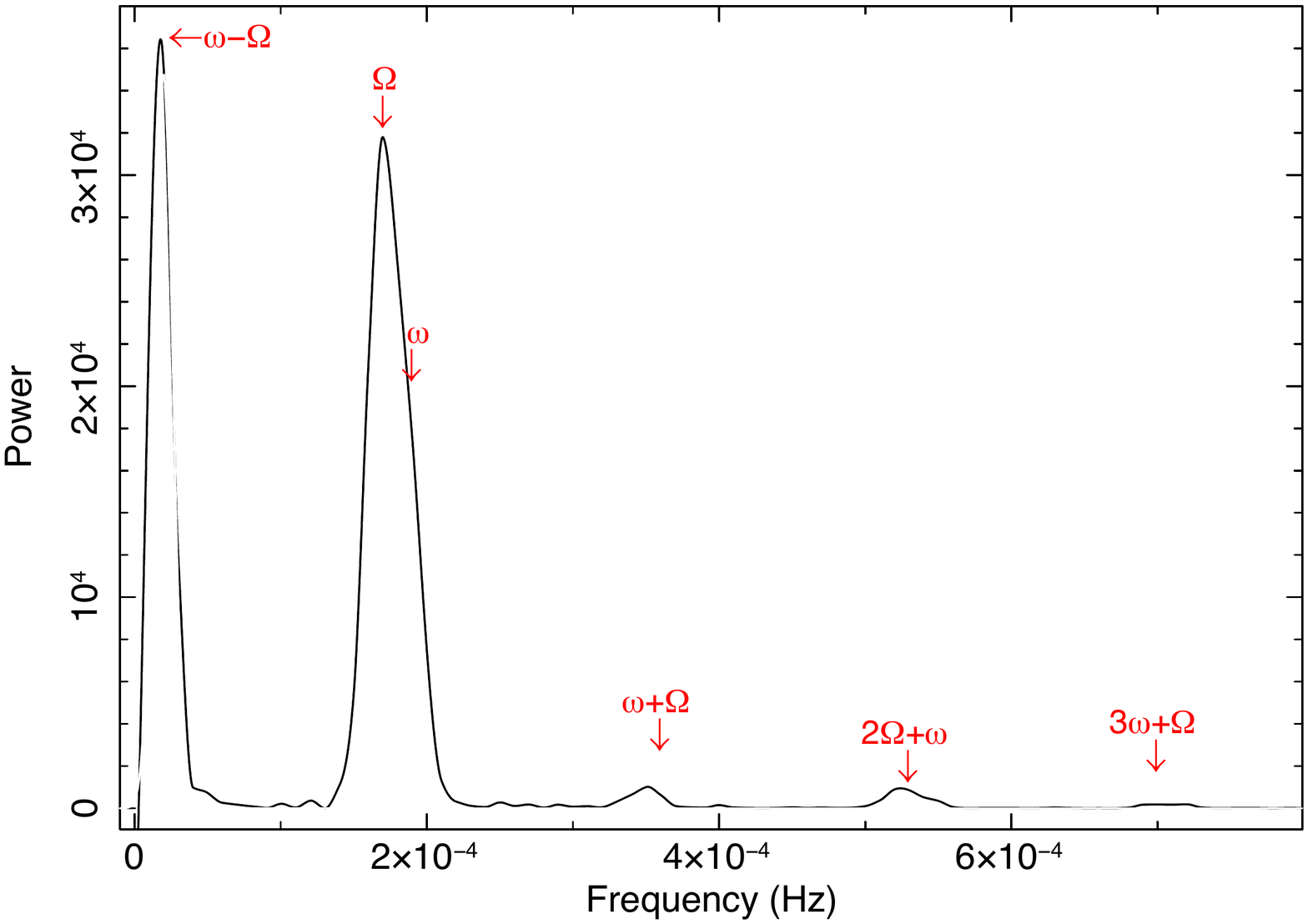}
\includegraphics[width=14cm,height=8.3cm]{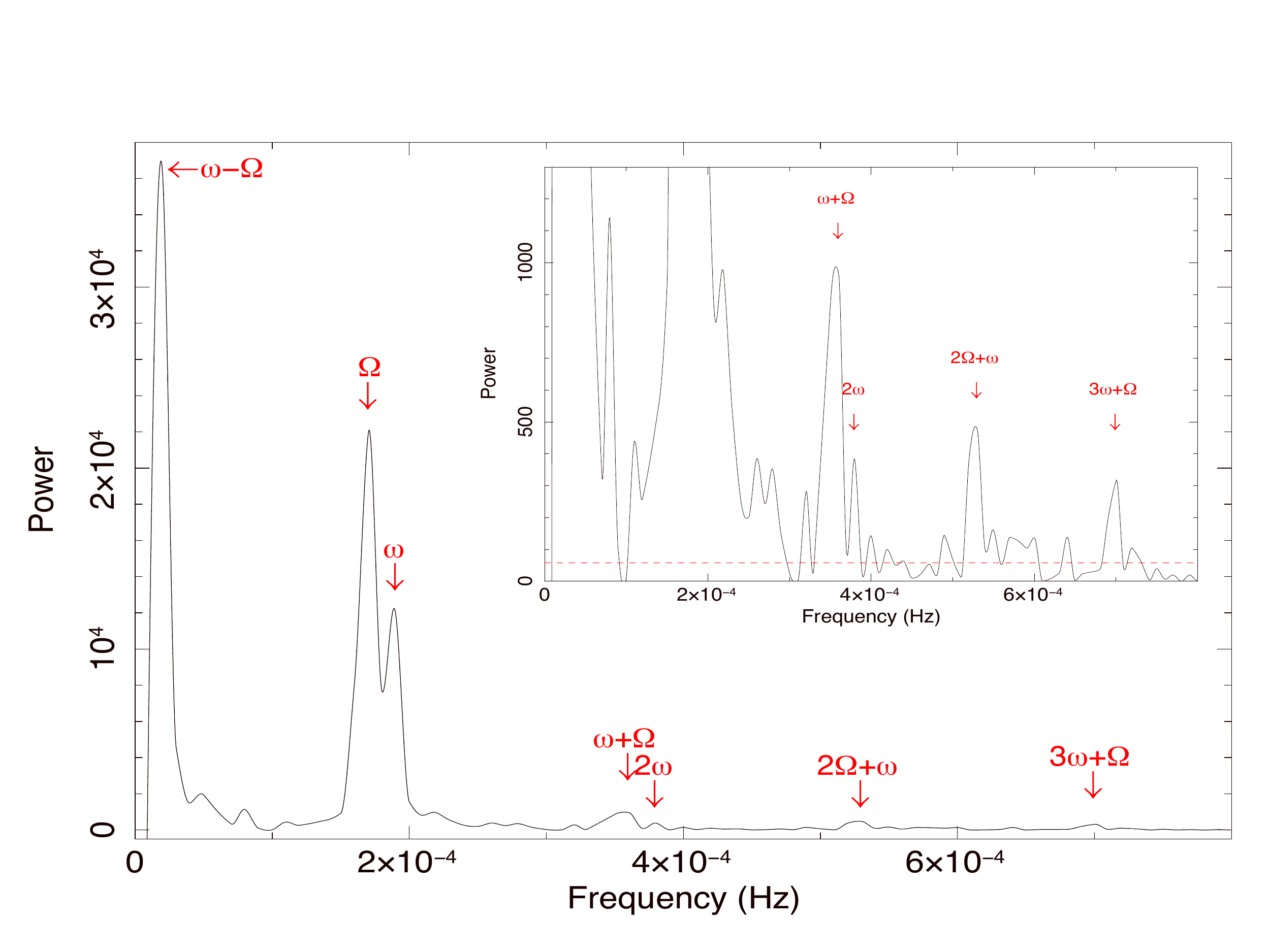}}
\caption{\label{fftxmm2} Power spectral distribution of \rx\ computed from the pn data of the first (left) and the second (right) \xmm\ observations. The frequencies/peaks corresponding to the 1.64\.h and 1.47\,h  modulations of the light curves are identified by $\Omega$ and $\omega$, respectively, and several other significant peaks can be interpreted as sidebands or higher harmonics of these frequencies (see sect.\,\ref{timing}). The inset is a zoom of the y-axis for clarity purposes, where the 5$\sigma$ significance threshold is reported with a dashed line.}  
\end{figure*}

The light curves of the two \xmm\ observations of \rx\ reveal substantial variations with time (see Fig.\,\ref{xmmlc} \,\ref{fftxmm2} and \ref{efolds}). A strong modulation with period of $\approx$1.6~h (6\,ks) is easily recognisable, with another ample modulation on a time scale approximately ten times longer. Since the longer modulation is present in both the \xmm\ observations, it might be recurrent as well, with a characteristic time $\approx$15\,h (54~ks). 

The modulation pattern is rather complex and we lack a prior knowledge of its origin, so we adopted the following approach to model it: we first fit to the data of each observation independently two sinusoidal function. Since the fit was not satisfactory, we tested the addition of further sinusoidal components with periods free to assume any value (i.e., we did not forced them to be integer factors or multiples of $P_1$ or $P_2$). In other words, we modelled the data the data with a Fourier sum of the form 
\begin{equation}A_0 + \sum_{k\geqslant2} A_k \sin \Bigg( \frac{2\pi(x-\phi_k)}{P_k}\Bigg).\end{equation}\label{function}
Of course, as $k$ increases, the adherence of the model to the data improves, but we found that the simplest function that fits reasonably well the data, adequately describing the occurrence of minima and maxima, and the overall shape of the curve, requires three sin components ($k=3$). By fitting this purely phenomenological model to the data, we derived the values $P_{1,\,\mathrm{XMM1}} = 1.644\pm0.005$\,h ($5.92\pm0.02$~ks), $P_{2,\,\mathrm{XMM1}} = 15.0\pm0.5$\,h ($54.2\pm2.0$~ks), and $P_{3,\,\mathrm{XMM1}} = 1.491\pm0.008$\,h ($5.37\pm0.03$~ks) for the first \xmm\ observation, and $P_{1,\,\mathrm{XMM2}} = 1.650\pm0.003$\,h ($5.94\pm0.01$~ks), $P_{2,\,\mathrm{XMM2}} = 15.2\pm0.2$\,h ($54.7\pm0.7$~ks), and $P_{3,\,\mathrm{XMM2}} = 1.477\pm0.003$\,h ($5.34\pm0.01$~ks) for the second. 
The measures of $P_1$ and $P_2$ are not precise enough to bootstrap a unique coherent solution that links the two \xmm\ data sets (in particular, there is a strong alias degeneracy in $P_1$).

To better assess the situation, we inspected a Fourier transform of the second \xmm\ observation computed using the pn data which allow the highest timing resolution (bin time: 47.66~ms). Several peaks can be observed in the power spectral density (PSD) plot (Fig.\,\ref{fftxmm2}). Two strong close but distinct peaks (their separation is larger than the intrinsic Fourier resolution, 
$\mathbf{\Delta\nu\simeq1.4\times10^{-5}}$~Hz) are located at $\Omega = 1.69(5)\times10^{-4}$~Hz (the central frequency was measured by the fit with a Gaussian, and is equivalent to $1.64\pm0.04$~h) and $\omega = 1.88(9)\times10^{-4}$~Hz ($1.47\pm0.08$~h), and they clearly correspond to $P_{1,\,\mathrm{XMM2}}$ and $P_{3,\,\mathrm{XMM2}}$, respectively. An even higher peak is present at $\sim$$1.87\times10^{-5}$~Hz (14.88~h) and is consistent with $P_{2,\,\mathrm{XMM2}}$. The fact that the central frequency of this peak differs by less than 2\% from the quantity $\omega$--$\Omega$, strongly suggests that it is the sideband (beat) frequency between $\Omega$ and $\omega$, 
which we then interpret as the orbital and spin period of \rx, respectively (see Sect.\,7.1).
Many other peaks that stand out above the significance threshold in the PSD can be interpreted as combinations of $\Omega$ and $\omega$. The most prominent ones are labelled in Fig.\,\ref{fftxmm2}. The PSD computed from the first observation is similar (Fig.\,\ref{fftxmm2}), but only one peak is present around the frequencies of $\Omega$ and $\omega$. This is to be expected in any case, because of the lower intrinsic resolution of this observation ($\sim$$\mathbf{2.2\times10^{-5}}$~Hz), but the slightly asymmetric shape of the peak, which is unbalanced towards $\Omega$, suggests that it may actually be a blend of two peaks.

\subsection{X-ray spectral analysis}
\label{spectrum}

We started the spectral analysis by fitting together the pn averaged-spectra of the two observations 
to a set of different models: power laws, bremsstrahlung, 
blackbodies, an accretion disc consisting of multiple blackbody components 
(\textsc{diskpbb}), a model describing the emission produced by a thermal distribution of electrons which
Compton up-scatter soft seed X-ray photons (\textsc{nthcomp}; Zdziarski, Johnson \& Magdziarz 1996; 
Zycki, Done \& Smith 1999), and a model reproducing the emission by hot diffuse gas (\textsc{mekal}).
We accounted for photoelectric absorption by the interstellar medium along the line of sight through the \textsc{tbabs} model, and 
adopted the photoionisation cross-sections from Verner et al. (1996) and the chemical abundances 
from Wilms, Allen \& McCray (2000). None of the above-mentioned models gave a satisfactory 
description of the data. 

Much more acceptable results were achieved when fitting the data to models consisting of several \textsc{mekal} components at different temperatures. 

\begin{figure*}
\centering
\hbox{
\includegraphics[width=10cm,height=10cm,angle=0]{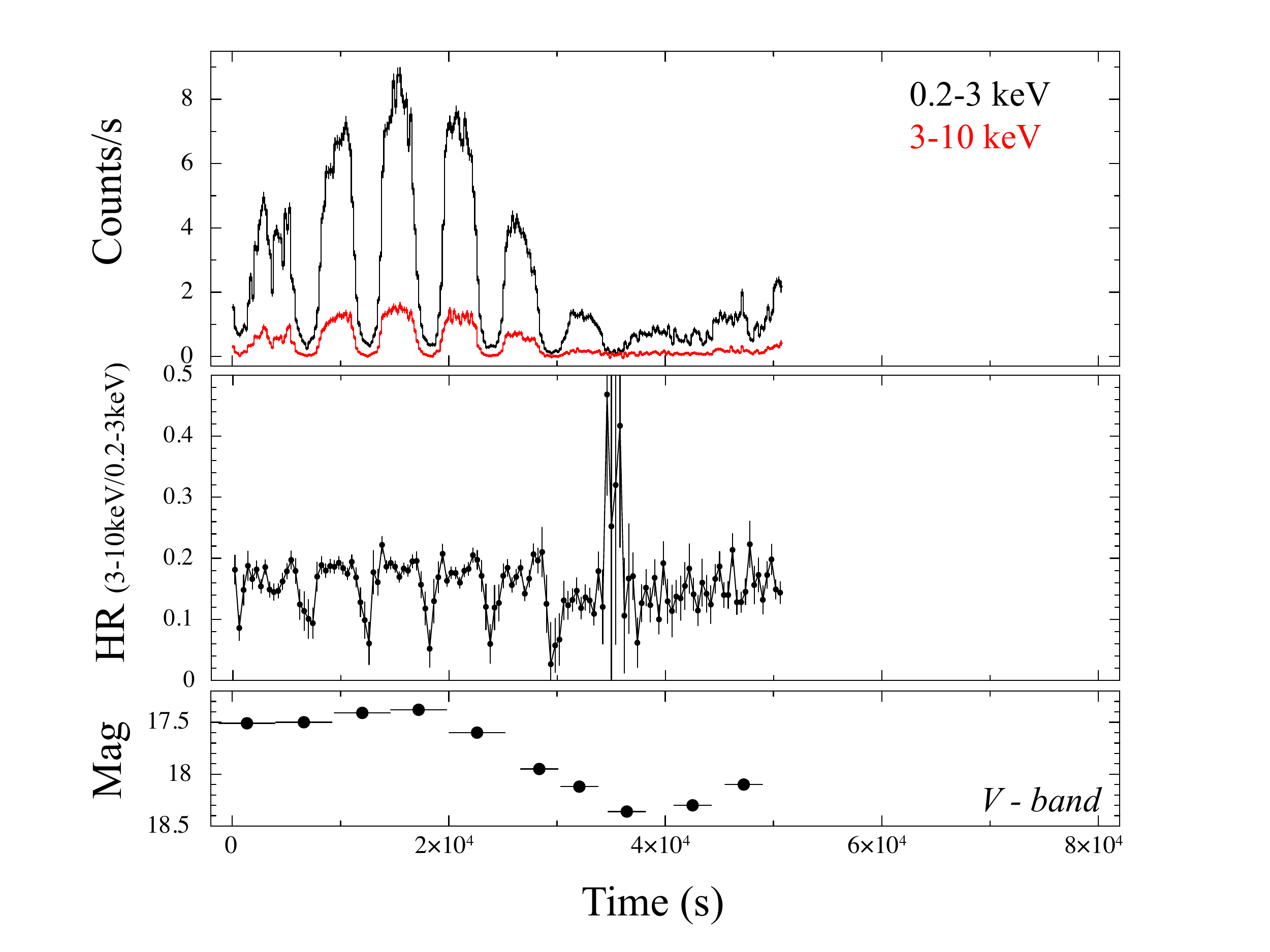}
\hspace{-1.3cm}
\includegraphics[width=10cm,height=10cm,angle=0]{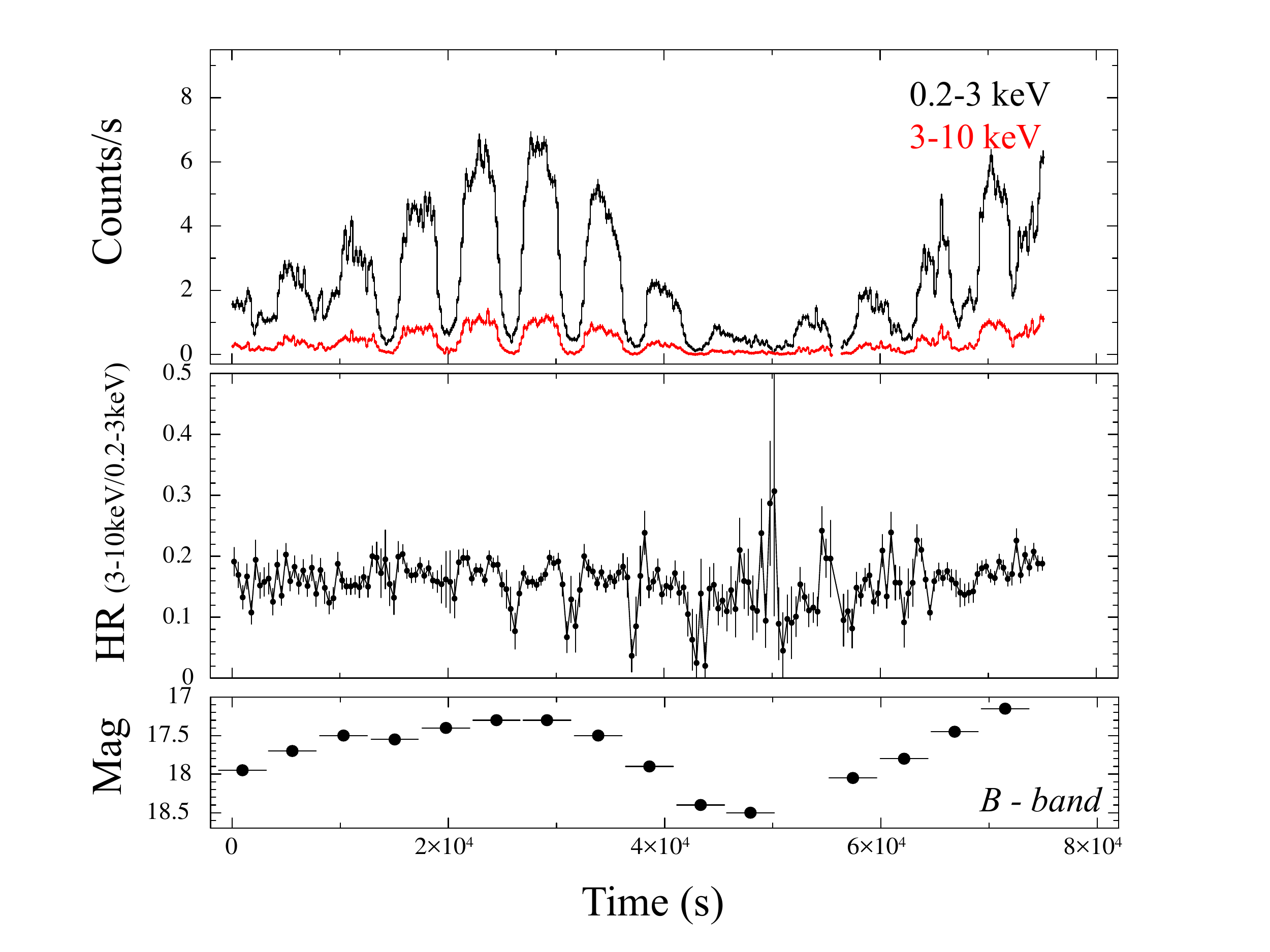}}
\caption{\xmm\, EPIC (pn+MOS) lightcurves (with a binning time of 200\,s) of \rx\, in two energy bands (and hardness ratios) for the first (left) and second (right) \xmm\, observation. In the bottom panels we report on the \xmm\, Optical Monitor magnitudes during the two observations (see text for details). Errors are smaller than the markers ($\sim$0.02--0.04 mag).}
\label{xmmlc}
\end{figure*}
\begin{figure*}
\begin{center}
\hbox{
\includegraphics[width=8cm]{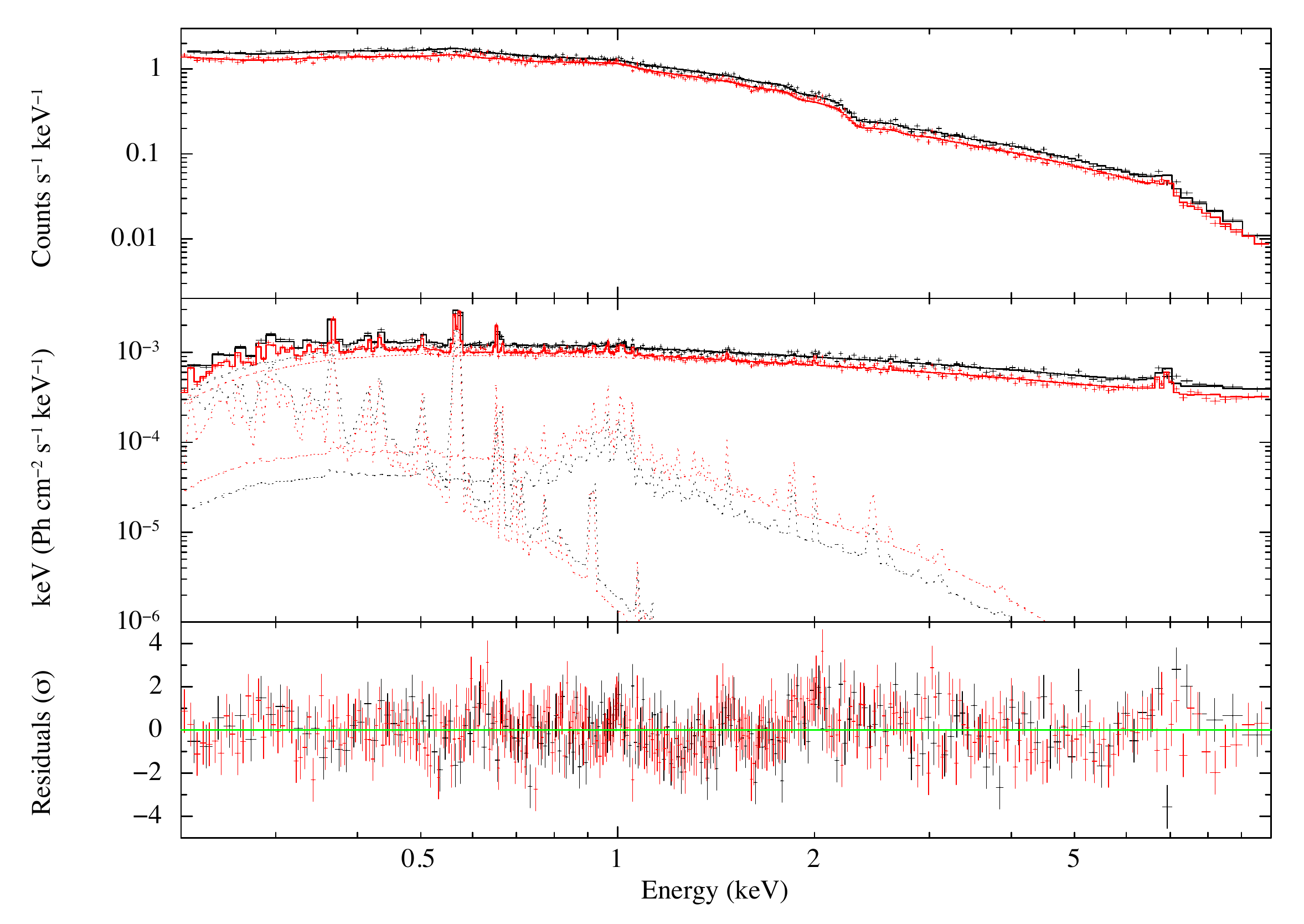}
\includegraphics[width=8cm]{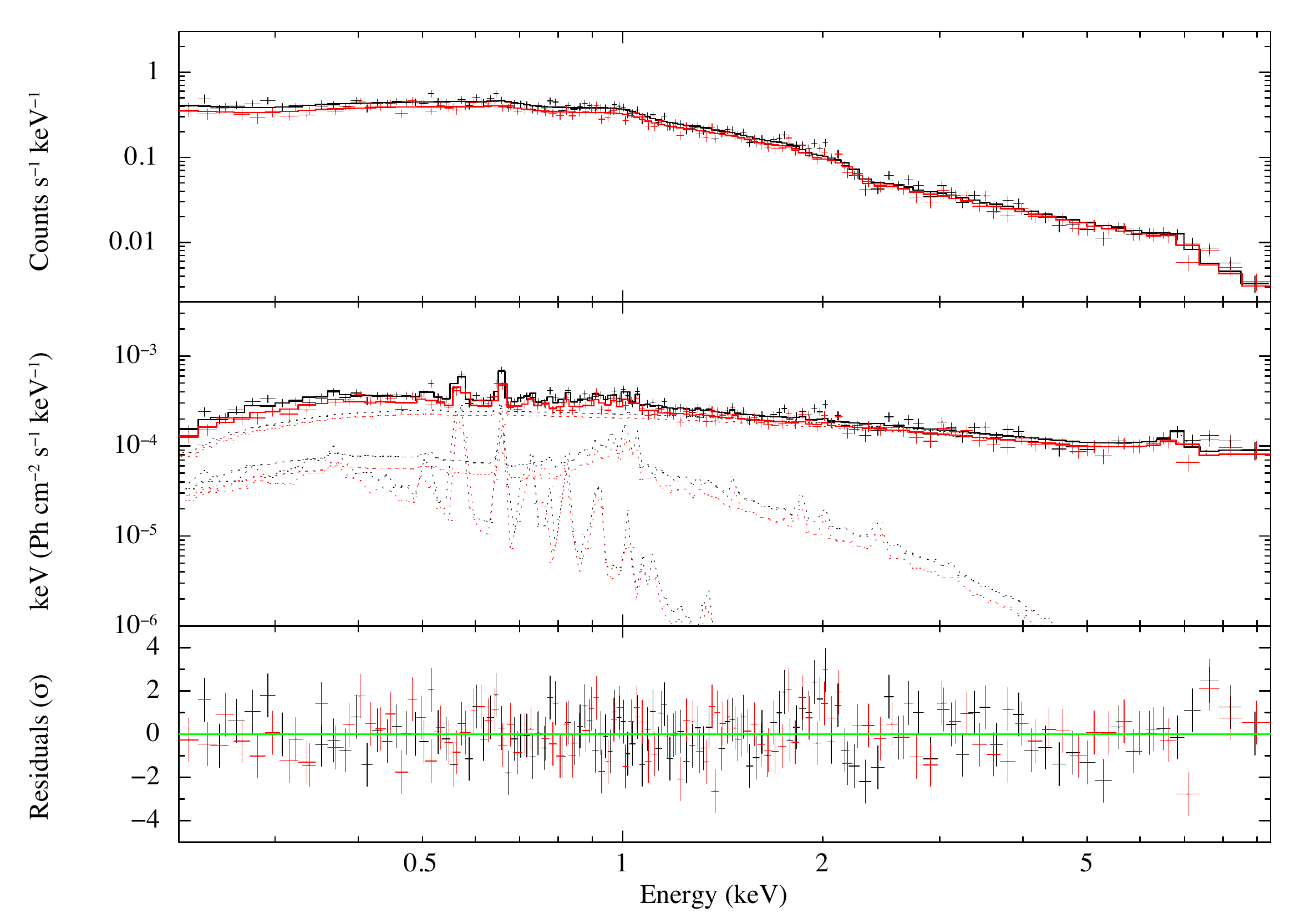}}
\caption{Left panels show the results of the spectral analysis of the peak spectra of the long-term modulation, while on the right we show the results of the minima. From top to bottom (in both panels): 0.2--10~keV pn spectra of \rx\ fitted to
the \textsc{tbabs*pcfabs*(mekal+mekal+mekal)} model (solid lines);  $E\times f(E)$ unfolded spectra; post-fit residuals in units of standard deviations. Black color 
refers to the first observation, red color to the second observation.}
\label{xmmspectra}
\end{center}
\end{figure*}

Given the strong flux variability in both observations, as well as the variable hardness ratio we observed comparing light-curves in different energy bands (see Fig.\,\ref{xmmlc}), we performed the spectral analysis comparing time-resolved spectra, which we believe yielding more meaningful results for this variable source (see Fig.\,\ref{xmmspectra}). We first analysed the spectra during the peak of the 54.7\,ks beat modulation for both observations, and then we further extracted spectra for the maximum and minimum of the shorter 5.9\,ks periodicity. Fitting with a series of 3\textsc{mekal} models we find good results for all the spectra we considered. Despite cycle-to-cycle variations observed in the hardness ratio in both observations (see Fig.\,\ref{xmmlc}), we did not find significant variability (within the 90\% parameter uncertainties) when fitting the 1.64\,h  phase-dependant spectra in any of the two observations nor merging them in phase (most probably due to the low number of counts we could collect at the minimum phases). However, we find significant changes in the spectra collected at the maximum and minimum of the long-term beat period of 54.7\,ks. Best-fitting parameters for the absorbed 3 \textsc{mekal} model are listed in Table~\ref{tab:spectral_parameters},  whereas the spectrum and the best-fitting model are shown in Fig.\,\ref{xmmspectra}. It is evident how the relative significance of the three different hot plasmas changes along the beat modulation, with the coolest component at an effective temperature of $\sim$0.2\,keV being more prominent at the maxima of the beat ($\omega$--$\Omega$) modulation. This might be due to accretion switching to the other (lower) pole-switching at half beat cycle, and the other pole possibly not totally visible (depending on the colatitude and the inclination angles). 

The source absorbed 0.2--10~keV flux during the first observation was $7.6\times10^{-12}$ \flux,  
a factor of $\sim$1.2 larger that measured during the second observation, about 1.5 months later. 
The  derived absorption column density, $\nh \sim1.6\times 10^{20}$~cm$^{-2}$, is about one order 
of magnitude lower than the estimated total Galactic value in the direction of the source ($\sim$$2\times 
10^{21}$~cm$^{-2}$; Willingale et al. 2013), implying a close-by location of this source within our Galaxy. 

\subsection{High resolution X-ray spectra}
\label{rgs}

We used the first order RGS spectra of both observations to search for narrow features. Two prominent narrow emission lines are clearly detected in both observations, plus a few weaker lines were also significantly detected. All the observed lines were well fitted by the multiple hot ionised plasma models (\textsc{mekal}) used to fit the EPIC-pn spectra (see Sec\,\ref{spectrum}). In particular, we identified in the RGS spectrum of the second XMM observations the lines corresponding to: Mg XII H-like Lyman $\alpha$ at 8.42 \AA, Ne X H-like Lyman $\alpha$ at 12.13 \AA, Ne IX He-like w-resonance at 13.45 \AA, Oxygen VIII H-like Lyman $\alpha$ (18.97 \AA) and $\beta$ (16.01 \AA), and triplet of the He-like Oxygen VII centred at 21.60 \AA\ (see Fig.\,\ref{fig:rgs}).
 
To test possible signs of variability among the most prominent lines between the maximum and minimum of the 1.64\,h X-ray modulation, we fitted the two phase-resolved RGS data (merging data of both observations) with an absorbed power-law plus two Gaussian functions with energies fixed at the 21.60\,\AA\ (0.57\,keV) and 18.97\,\AA\ (0.65\,keV), and keeping all spectral parameters forced to be the same except for the normalizations. The equivalent width of the Oxygen VII line is 64$^{+34}_{-17}$\,eV in the RGS spectrum at the maximum of the modulation, while only an upper limit of $< 62$\,eV (at 90\%) could be derived in the RGS spectrum at the minimum of the 1.64\,h modulation. Similarly, the equivalent width of the Oxygen VIII line is 33$^{+6}_{-13}$\,eV in the RGS spectrum at the maximum, and we derived an upper limit of $<$69\,eV at the minimum.  No significant phase-dependent variability is detected in the emission lines, and the Oxygen lines are well reproduce by the lowest temperature ($\sim$0.2keV) MEKAL component.


\subsection{Long-term X-ray light curve}

The long-term light curve of \rx\ presented in Fig.\,\ref{fig:xrt_lcurve} was extracted after fitting all \swift/XRT spectra with an absorbed power law model with the absorption column density fixed to the value derived from the fits of the \xmm\ data (the lower counting statistics of the XRT data precluded a more detailed modeling as done for the EPIC data sets). We obtained $\chi^2_\nu = 1.13$ for 203 dof. We deem that the large scatter in the flux values clearly visible in the figure (especially in the most recent observations) are likely due to the fact that the short exposures of the observations (see Table~\ref{tab:xray_log}) have sampled the source $\approx 54$-ks cycle at random phases. However, the long-term light-curve indicates that no substantial variation is observed in the source X-ray flux across the past 6 years.

\subsection{Results from optical observations}

The optical spectra of \rx\, (see Fig.\,\ref{nttspectrum}) are dominated by prominent emission lines such as H$\beta$, He I ($\lambda$ 4471, 4919, 5015), He II ($\lambda$ 4686) and possibly bowen-blend N III ($\lambda$ 4634-41)/C III ($\lambda$ 4647-50). No significant absorption feature is detected, in agreement with the findings of Masetti et al. (2013). 

\begin{figure}
\includegraphics[width=8cm,angle=0]{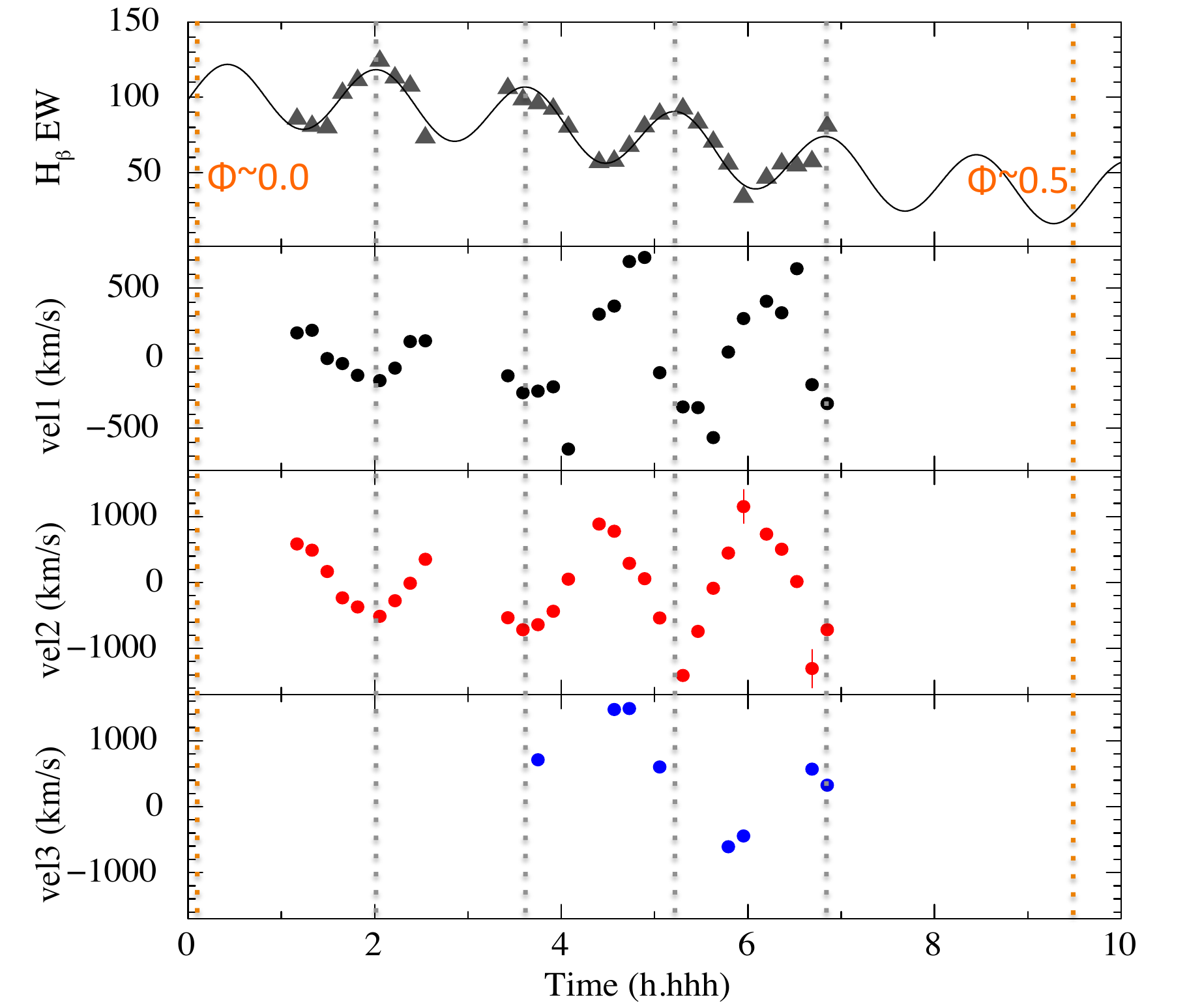}
\caption{\label{velocities} Equivalent width of the H$\beta$ line (top panel),  and velocities of three H$\beta$ components as a function of time as observed from ESO-NTT. Vertical dashed-lines report on the phase 0 and 0.5 of the 15.2\,h modulation (orange), and on the maxima of the 1.64\,h orbital modulation (grey).}
\end{figure}

Among the emission features detected, H$\beta$ is the most prominent. In the individual spectra we found no evidence for double-horned profile (usually interpreted as a signature for an accretion disc). The line profile is however clearly asymmetric. We tried to model the H$\beta$ line with a set of multiple Gaussians and found a reasonable solution using three Gaussians plus a constant: in Fig.\,\ref{velocities} we show the velocities of the these three components. The H$\beta$ line consists of a narrow (FWHM =11 \AA\ $\sim$ 680\,km s$^{-1}$) a broad (FWHM $\sim$ 30 \AA\ $\sim$ 1900 km s$^{-1}$) and a very narrow (FWHM $\sim$ 6 \AA\ $\sim$ 370 km s$^{-1}$) components, the latter statistically detected only in a few spectra (see Fig.\,\ref{velocities}). The large width of the broad component indicates it probably forms in the magnetically confined accretion flow, the narrow one likely forming in the ballistic trajectory of the accretion stream and the very narrow component from the irradiated hemisphere of the donor star.  These component are generally observed in Polars and particularly in asynchronous polars (see eg. Schwarz et al. 2005).

\begin{figure}
\centering
\vbox{
\includegraphics[width=8cm]{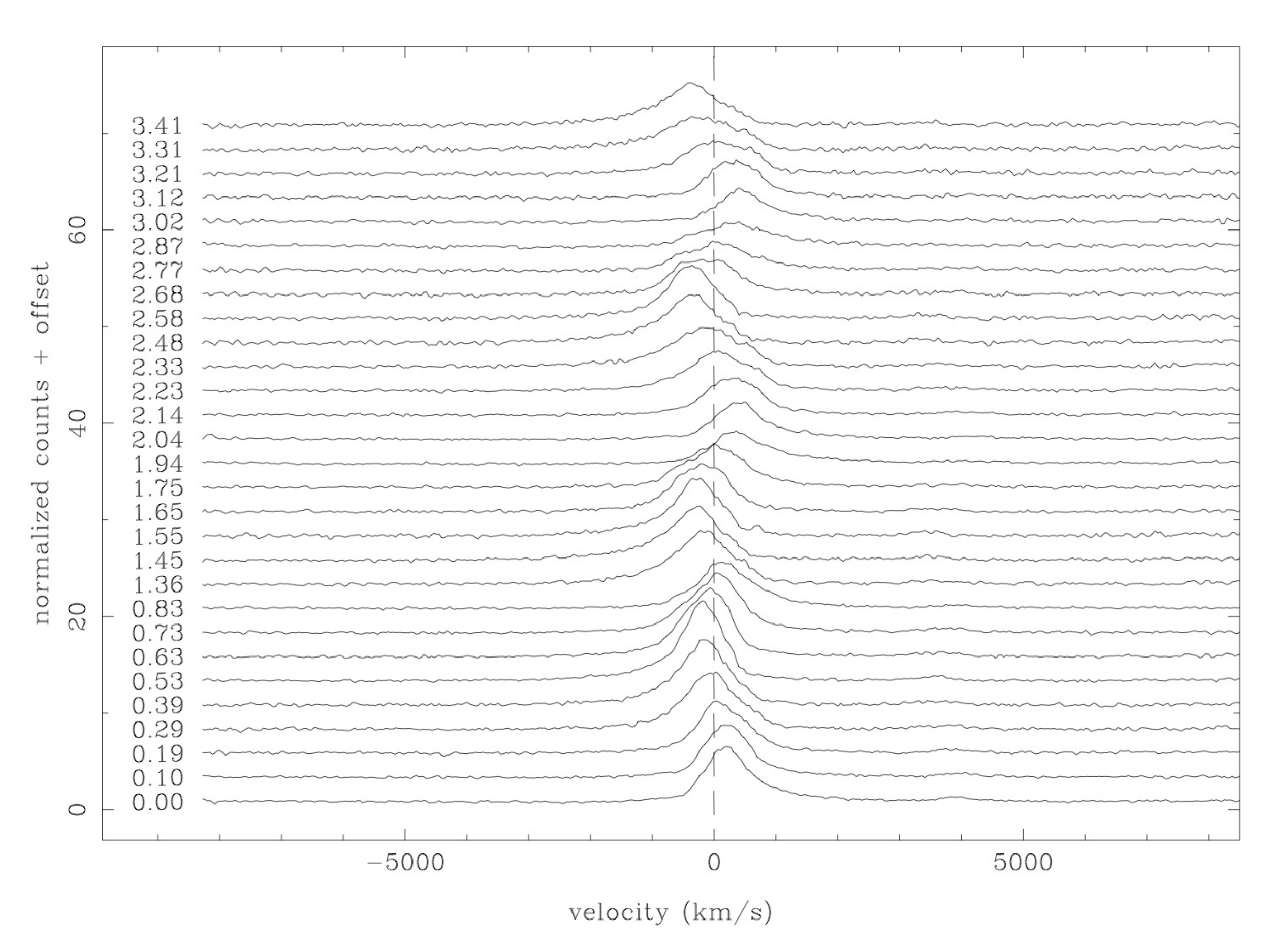}
\includegraphics[width=8cm]{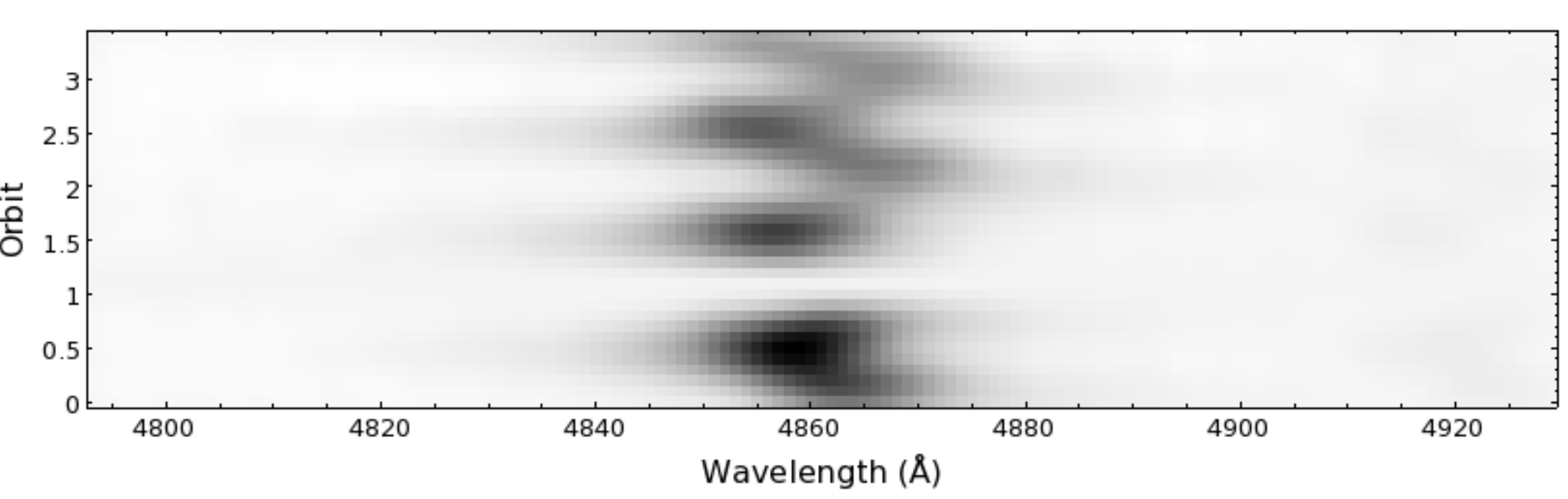}}
\caption{\label{ntt} ESO-NTT spectra of \rx. Top panel: The $H_{\beta}$ emission line as a function of the orbital phase (computed assuming phase zero for the first NTT spectrum). Bottom panel: Trailed spectra centered on the $H_{\beta}$ emission line; orbital phases are relative to the 1.64\,hr periodicity.}
\end{figure}

Fits to the radial velocity measurements for each lines have been first attempted by using a constant and a sinusoidal function (see Fig.\,\ref{velocities}).  We here use the radial velocities as derived for the H$\beta$ narrow and broad components, and the radial velocity measures of the other emission lines obtained through single-gaussian.
A fit with a constant plus a single sinusoidal function never provided a satisfactory result in terms of reduced $\chi^2$. We then tried a different approach and fit each radial velocity curve with the function of eq.(1) with $k=3$, $P_2 =P_{2,\,\mathrm{XMM2}} = 15.2$\,h and $P_3 =P_{3,\,\mathrm{XMM2}} = 1.47$\,h  (fixed); see Fig.\,\ref{velocities}).
We measured the following $P_1$ periods: $1.75\pm0.05$~h (H$\beta$, first velocity, $\chi^2_\nu/\mathrm{dof}=1.40/21$), $1.72\pm0.05$~h (H$\beta$, second component $\chi^2_\nu/\mathrm{dof}=0.71/21$), $1.53\pm0.01$~h (He\textsc{i} 4480~\AA, $\chi^2_\nu/\mathrm{dof}=0.67/20$), $1.48\pm0.01$~ks (He\textsc{i} 4920~\AA, $\chi^2_\nu/\mathrm{dof}=4.53/20$), and $1.66\pm0.04$~h (He\textsc{i} 5020~\AA, $\chi^2_\nu/\mathrm{dof}=0.61/20$). 
From the first H$\beta$ component, we were able to derive some constraints also on the amplitudes: $A_0 = -$$(4\pm1)\times10^2$ km s$^{-1}$, $A_1 = (4.2\pm0.5)\times10^2$ km s$^{-1}$, $A_2 = (6.3\pm1.6)\times10^2$ km s$^{-1}$, $A_3 = (1.7\pm0.3)\times10^2$ km s$^{-1}$. This multi-sinusoidal representation indicates that besides the orbital modulation (with amplitude of about 400km/s - see Fig.\,\ref{ntt}) a fast moving high velocity component shifting from a maximum of about $+1100$ to $-1200$\,km/s may be due to the ballistic accretion stream which changes also along the beat cycle.

For each spectrum, we also measured the equivalent width (EW) of the H$\beta$ and He I emission lines. We found evidence for significant EW variability, as observed also for the optical flux measured by our IAC-80 and \xmm/OM observations (see e.g. Fig.\,\ref{xmmlc} right panel-bottom). The EW of all emission features seems to be modulated over the 1.64\,h and 15.2\,h  periodicities, the orbital period and the longterm beat periodicity (see Fig.\,\ref{velocities}).   Since the spectral lines EW is a proxy of the source optical flux, such findings are not unexpected.

The NTT data consist of short exposures of $\sim$500~s each, spanning only a few cycles of the orbital modulation at 1.64\,h, for a total observation time of $\sim$6\,h (see Fig.\,\ref{ntt} and \ref{velocities}). For this reason, when fitting them with the multicomponent models used in Section\,\ref{xmmtiming}, we held $P_2$ and $P_3$ at the \xmm\ values measured during the second observation. 

Because of the very sparse sampling, long data gaps, and short exposures, the IAC80 photometric data are of limited help in constraining the model and/or refining the periods. So, we fixed $P_1 =P_{1,\,\mathrm{XMM2}} = 1.64$\,h and $P_3 =P_{3,\,\mathrm{XMM2}} = 1.47$~h (the 3-sin model provides also for the IAC80 data a much better fit than the 2-sin model). In this way, we found for the long modulation $P_{2,\,\mathrm{B}}\simeq15.1$~h from the $B$-filter data and $P_{2,\,\mathrm{R}}\simeq15.2$~h from the $R$-filter data.


\section{The other two X-ray sources in the field: \tmsp\ and \thirdxmm}
\label{others}

Of the two additional X-ray sources within the field of the mCV, \tmsp\ can be identified in the USNO A2 
Catalogue at $\rm RA=08^h38^m50\fs4$ $\rm Dec=-28\degr27'56\farcs7$ (J2000.0) with $R=18.0$ and $B=20.6$, 
and \thirdxmm\ is catalogued in the Wide-field Infrared Survey Explorer (WISE\footnote{See http://wise2.ipac.caltech.edu/docs/release/allsky/.}; 
Wright et al. 2010) as WISE\,J083842.77$-$282830.9. Its position is $\rm RA=08^h38^m42\fs777$, $\rm Dec= -28\degr28'30\farcs99$ (J2000.0), 
and it has optical magnitudes in the USNO B1 Catalog of $B1\sim19.66$, $B2\sim18.49$, $R2\sim19.02$ and $I\sim18.01$. 

We carried out a detailed X-ray spectral and timing analysis for both sources. \tmsp\ appears to be rather variable in the X-rays (see the values for the root mean square fractional 
variation and fluxes in Table~\ref{tab:log_sources}), and showed a $\approx 600$-s long flare with a structured morphology in the second observation (see Fig.\,\ref{src2_flare}).
\footnote{A careful analysis of the background light curve excludes a background 
flare origin}. For both sources, spectra of the two observations were fitted together with an absorbed power law model, and the absorption 
column density was tied up across the data sets (for the case of \tmsp\ both the flare and quiescent spectra were extracted).  Best-fitting parameters are reported in Table~\ref{tab:log_sources}. The steady emission of \tmsp\ is brighter during the second observation by a factor of $\sim$5, while showing flaring episode, and its spectrum slightly harder than in the first observations. On the other hand, \thirdxmm\ appears to be steady during the X-ray observations.


\begin{table*}
\tiny
\centering 
\caption{Positions, pn count rates and spectral fit results for the X-ray sources within the error box of 3FGL\,J0838.8$-$2829. The source numbering is as 
shown in Fig.\,\ref{src2_flare}. The pn spectra were fitted to the \textsc{tbabs*pegpwrlw} model in the 0.2--10~keV energy range. Uncertainties 
are quoted at a 90\% confidence level for a single parameter of interest, whereas upper limits are given at the 3$\sigma$ confidence level. (q) and (f) refer to the quiescent and
flaring states of \tmsp, respectively, during the second observation.}
\label{tab:log_sources}
\begin{tabular}{@{}cccccccccc}
\hline
Source			& Obs ID & $\rm RA$						& $\rm Dec$						& Count rate$^a$ 			& rms$^b$ 				& $\nh$					& $\Gamma$ 					& Unabs flux$^a$ 			& $\chi^2_\nu$ (dof)	\\ 	  	
			  	&  & \multicolumn{2}{c}{(J2000.0)}											& ($\times 10^{-3}$ s$^{-1}$)	&	  					& ($10^{21}$ cm$^{-2}$)		&							& ($10^{-14}$ \flux)			&		\\
\hline
\multirow{3}{*}{\tmsp}	& 0764420101  & \multirow{3}{*}{$08^h38^m50\fs40$}	& \multirow{3}{*}{$-28^\circ27'59\farcs04$}& $16.7 \pm 0.9$			 	& $0.37 \pm 0.06$			&  $0.3\pm0.2$	& $1.6\pm0.2$		& $8\pm1$				& \multirow{3}{*}{0.97 (63)}\\
				&  \multirow{2}{*}{0790180101}  &								&								& $74 \pm 2$ (q) 	& \multirow{2}{*}{$1.23 \pm 0.07$}	& tied						& $1.15\pm0.08$ (q) 	& $38 \pm 3$	(q) 				& \\										&			 &								&								& $1223 \pm 82$ (f)	 		& 				& tied						& $1.3\pm0.2$ (f)			&	$550 \pm 90$ (f)		& \\	
\hline
\multirow{2}{*}{\thirdxmm}	& 0764420101 & \multirow{2}{*}{$08^h38^m42\fs72$}	&  \multirow{2}{*}{$-28^\circ28'32\farcs52$} & $22.1 \pm 0.9$ 			& $< 0.18$		& $1.1\pm0.3$	& $1.9\pm0.2$		& $11.1\pm0.9$			& \multirow{2}{*}{0.94 (27)} \\
				& 0790180101 &								&								& $14.8 \pm 0.8$			& $< 0.17$		& tied						& $2.1\pm0.2$							& $8.6\pm0.9$ 		& \\
\hline						
\end{tabular}
\begin{list}{}{}
\item[$^{a}$] In the 0.2--10~keV energy range.
\item[$^{b}$] Root mean square fractional variation, determined by applying the \textsc{lcstats} (v. 1.0) tool of \textsc{xronos} on the 0.2--10~keV background-subtracted 
and exposure-corrected light curves from the pn camera with a binning time of 400~s.
\end{list}
\end{table*}

\section{Discussion}
\label{discussion}

Our multi-band study of the field of \fermirx\, revealed two variable X-ray sources: \rx\, and \tmsp. We discuss here our results and the nature of these two intriguing objects.

\subsection{\rx: an asynchronous magnetic CV}

\begin{figure}
\includegraphics[width=1.1\columnwidth,angle=0]{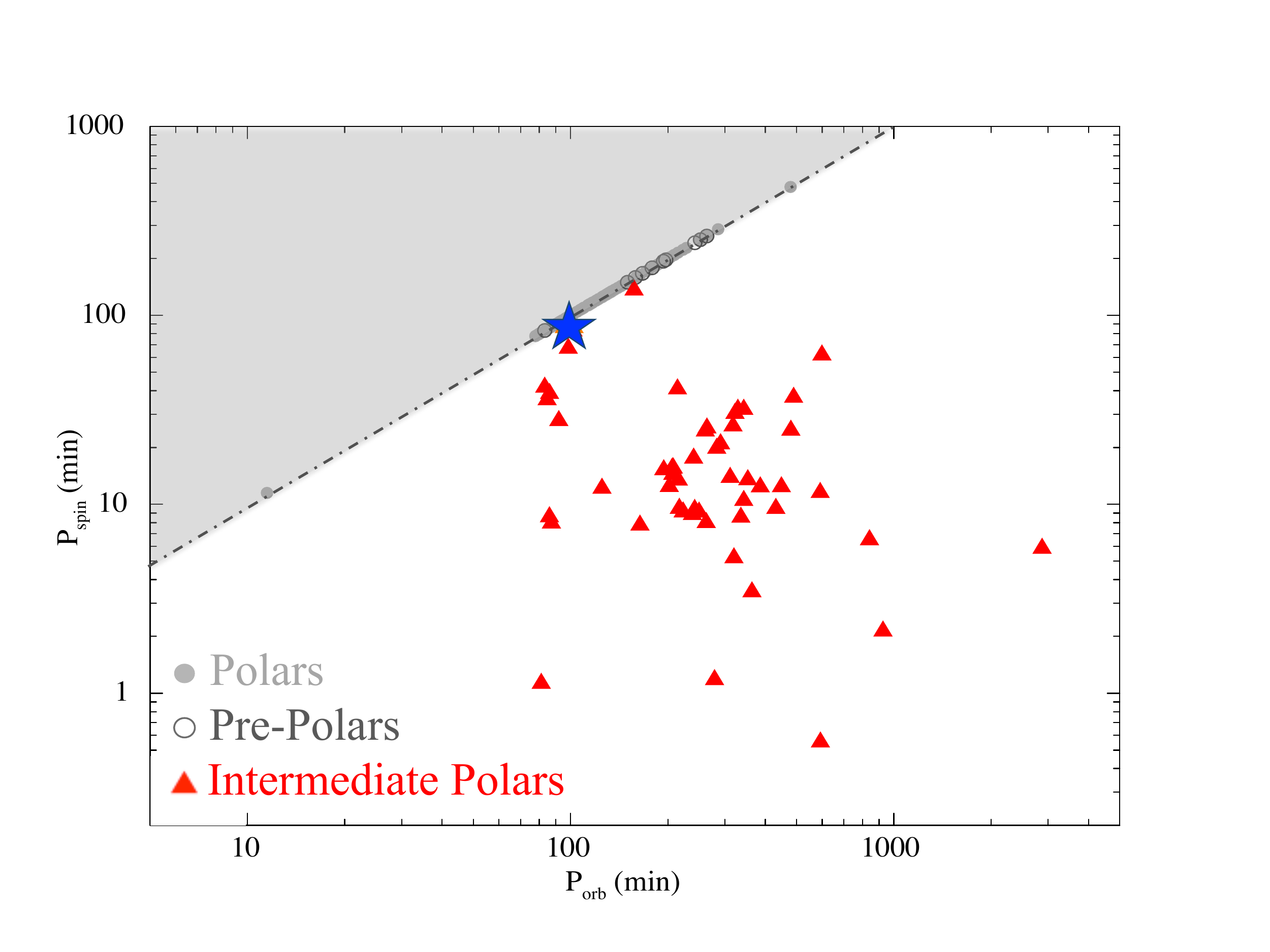}
\vspace{-0.5cm}
\caption{Orbital and spin periods of polars (grey dots), pre-polars (dark grey circles), and IPs (red triangles) known to date, with the values derived for \rx\, shown as a blue star. The dot-dashed line represent $P_{\rm spin} = P_{\rm orb}$. Data adapted from Ferrario, de Martino \& G\"ansicke (2015). }
\label{periods}
\end{figure}

\rx\, has shown to be highly variable on short and long timescales with a complex behaviour. In particular, we detect two dominant periodicities at 1.64\,h and 15.2\,h in the power spectrum as well 
as a 1.47\,h peak which is about half the power of the 1.64\,h one (see Fig.\,\ref{fftxmm2}). These are remarkably correlated
as the long periodicity is found precisely at the beat between the two shorter ones. These characteristics and their
timescales are uncommon to other variable X-ray sources, as the LMXBs, but are consistent with those observed in CVs of the 
magnetic type. 
The close 1.64\,h and 1.47\,h periods if interpreted as the binary orbital period and the rotational period of the WD give $P_{\mathrm{spin}}/P_{\mathrm{orb}}\sim0.90$. The location of \rx\, in the 
spin--orbit period plane of mCVs is shown in Fig.\,\ref{periods}, and strongly suggests this system is 
close to, but not right at, synchronism. 
We therefore have identified either one of the slowest rotating IP (see also Coti Zelati 
et al. 2016) or a highly de-synchronised polar. Other weakly asynchronous 
magnetic systems with similar spin-orbit ratios are Paloma, $P_{\mathrm{spin}}/P_{\mathrm{orb}}\sim$ 0.93 (Schwarz 
et al. 2007), and IGR\,J19552+0044 with $P_{\mathrm{spin}}/P_{\mathrm{orb}}\sim$ 0.82 (Bernardini 
et al. 2013). The other four confirmed asynchronous polars are only 
very slightly de-synchronised ($P_{\mathrm{spin}}/P_{\mathrm{orb}} > 0.98$; see also Fig.\,\ref{tab:mcv_properties}).

\begin{figure}
\includegraphics[width=8.5cm,angle=0]{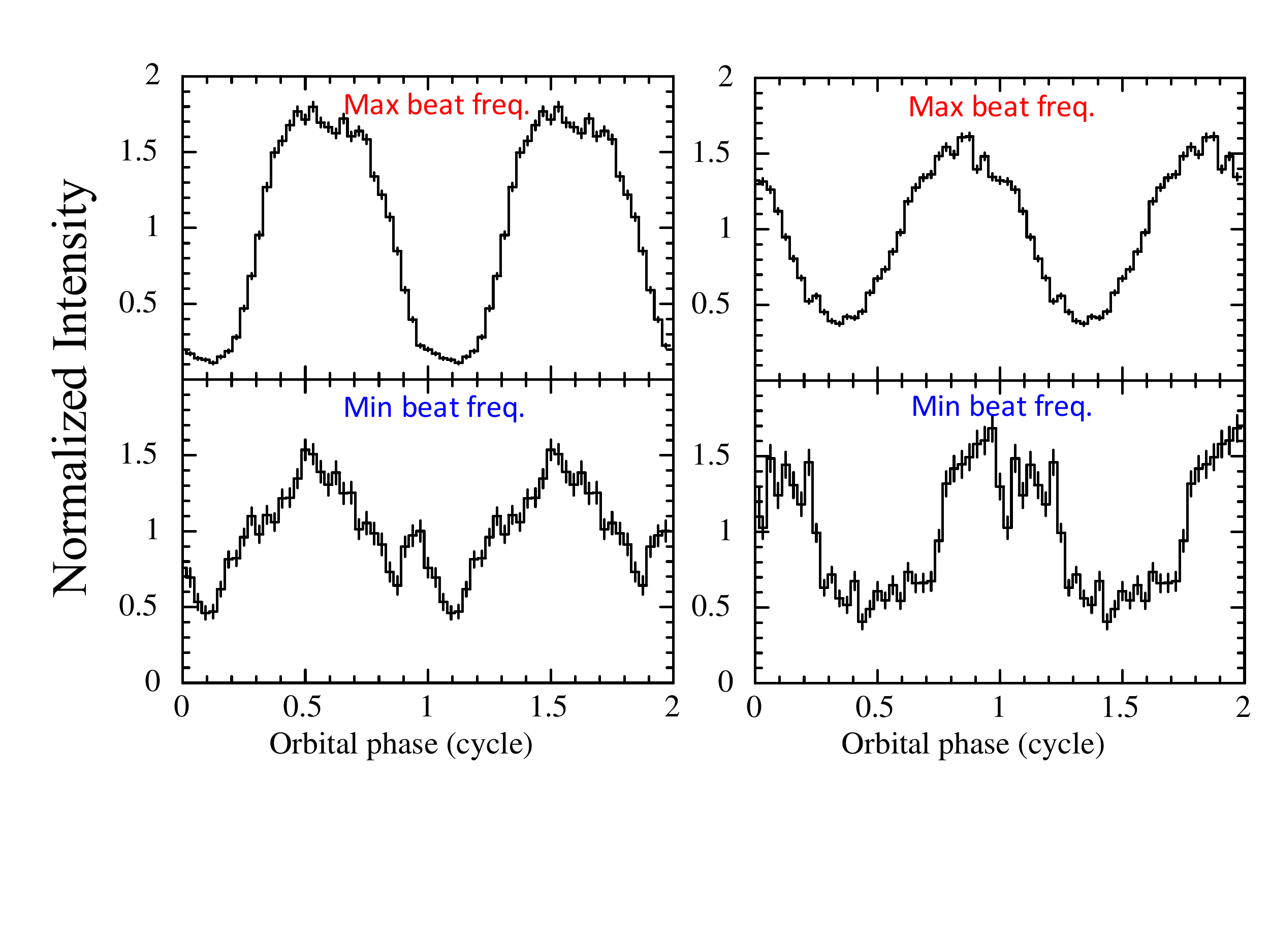}
\vspace{-1cm}
\caption{\xmm\, light-curves for the first (left) and second (right) observations folded at the system orbital period. Top panels are relative to the time-span of the maximum of the beat modulation, and bottom panels to the minimum.}
\label{efolds}
\end{figure}

Folding the X-ray light curve at the 1.64\,h period (see Fig.\,\ref{efolds}; or at the 1.47\,h too) the shape is almost flat-topped closely resembling those observed 
in the strongly magnetic polar systems (Cropper 1990), where the modulation is due to 
self-occultation of the accretion pole as the WD rotates. The spectrum is thermal, requiring 
three optically thin emitting regions with temperatures 0.2\,keV, 1\,keV and 14\,keV highly
absorbed by dense partial ($\sim$70\%) covering material ($\sim$$2\times10^{24}$\,cm$^{-2}$). The
spectral characteristics also closely resemble those observed in magnetic CVs, both polars
and IPs (e.g. Ramsay \& Cropper 2004; Bernardini et al. 2012). The 
lack of spectral changes (except for the normalization) at the $\sim$1.6--1.5\,h periods is consistent 
with the spectral behaviour seen in polars due to the self-occultation of the main 
accreting pole, while IPs instead show highly variable absorption along the spin period.
Furthermore, and differently from many polars, \rx\, does not
show any evidence of a soft optically thick (blackbody) component. This component
originates from reprocessing of hard X-rays and cyclotron radiation at the WD 
polar regions. 
The presence of this component has been for a long time considered ubiquitous in polars,
until new observations have shown that this is not necessarily so 
(Ramsay \& Cropper 2004). On the other hand, IPs have typically hard spectra, and  the
recent identification of a soft blackbody component in about 30\% of these systems
(Anzolin et al. 2008; Bernardini et al. 2012) does not make these characteristics
an identifying feature of one or the other type of mCVs.

The long observations acquired with XMM-Newton have allowed an unprecedented 
monitoring of \rx\, revealing a remarkable long-term 15.2\,h X-ray modulation which repeats
itself in the two observations taken 1.5 months apart that we naturally interpret as the
the beat period between the 1.64\,h and 1.47\,h periodicities. The spectrum changes mainly in the
normalizations of the optically thin components and in the temperature of the warmer plasma
but not in the partial absorption components. This further strengthens our interpretation that
this variability is not the orbital period since the orbital X-ray variability in the 
asynchronously rotating IPs is found to be due to absorbing material fixed in the binary 
frame such as the hot-spot at the rim of an accretion disc or eclipses in high inclination systems (see Parker et al. 2005). 
The optical radial velocities of emission lines and the optical light also give evidence 
for the presence of such long variability implying that it does not arise from X-ray irradiation on an accretion disc as seen in many IPs (see Warner 2003). 
The presence of this modulation in both X-rays and optical implies a diskless 
configuration (Wynn \& King 1992), in which the accreting 
material alternates onto the WD poles. In non-synchronous systems, the flow rotates
around the magnetic field of the WD on a time-scale of the beat period. This has
the effect that the flow will be preferentially directed onto the first (main) pole
and then to the other, each half beat cycle.
The weaker modulations at the 1.47\,h and 1.64\,h periods at about half the 15.2\,h beat period indicates that there is pole
switching in \rx. In particular diskless models predict the dominance of the beat and of the orbital frequencies and a much lower or even undetectable spin peak in the power spectrum in configurations of moderately high values of the binary inclination angle ($i$) and magnetic colatitude ($m$) provided that $i+m > 90^o + \beta$, where $\beta$ is the pole-cap opening angle  (Wynn \& King 1992). The presence of lower peaks at sidebands  depends on pole-switching effect that modulates the light curve.  Furthermore the power of the spin peak is dependent on the degree of asymmetry, with strong asymmetries producing weaker spin pulsations (Wynn \& King 1992). This appears to be the case in \rx. Additionally the 
slope of the beat modulation depends on the transition times of accretion flipping onto the two poles and thus the almost smooth variability at the 15.2\,h period would indeed indicate that the stream from the donor star does not switches between the two poles instantaneously. We 
furthermore note
that a large $\beta$ tends to produce either sinusoidal or flat-topped light 
curves indicating that accretion in \rx\, occurs over wide areas of the WD 
poles. Furthermore, the faint phase of the modulation at the 1.64\,h (or 1.47\,h) 
period (see Fig.\,\ref{efolds})  observed at beat maximum lasts $\sim$0.3. This 
can be used to estimate the ranges of binary
inclination $i$ and magnetic colatitude $m$ (Cropper 1990) as: 
$ cot\, i$ = $\cos (\pi \, \delta\phi ) \times \tan (m) $, where $\delta\phi$ is the
length of the faint phase. Also, the lack of X-ray eclipses restricts 
$i \lesssim $75$^o$. We then find $24^o \lesssim m \lesssim 59^o$ for 
$ 10^o \lesssim i \lesssim 75^o$. 
A detailed modelling of accretion region is however beyond the 
scope of this work and 
will be subject of a future publication. The flipping accretion
configuration are likely 
occurring also in other asynchronous systems such as such as in Paloma 
(Schwarz et al. 2007), BY Cam (Piirola et al. 1994) and CD Ind (Ramsay 
et al. 1999), and in V1432 Aql (Mukai et al. 2003). In the latter system, 
the hard X-ray emission is found to be dominated by the orbital variability 
along the beat cycle whilst the X-ray spin variability is almost undetected 
in a long \emph{BeppoSAX} observation (Mukai et al. 2003) and with complex 
structure in \xmm\, and {\it R}XTE observations (Rana et al. 2005). 

The short orbital period of \rx\, puts this mCV below the 2--3\,h orbital period gap of CVs, where most polars are found (Ferrario, de Martino \& Gaensicke 2015; see also Fig.\,\ref{periods}).  An orbital period of 1.64\,h (98\,min) is close to that of
the IP EX\,Hya which is also (but not so weakly) de-synchronised ($P_{\mathrm{spin}}\sim67$\,min). Its spin-orbit
period ratio ($P_{\mathrm{spin}}/P_{\mathrm{orb}}=0.68$), similar to the few IPs found below the gap, could indicate
that these systems will never reach synchronism (see Norton et al. 2004).
Looking at the other two weakly de-synchronised mCVs, Paloma and IGR\,J19552+0044,
they are both short period systems, the former with a 157\,min orbital period falls right inside 
the orbital period gap and the second with a 101\,min orbital period
is below the gap. They are likely in a transition stage before reaching synchronism and thus believed to be polar progenitor candidate. On the other hand, the four asynchronous polars are instead less de-synchronised. Synchronisation timescales of $\sim$170 yr and $\sim$200 yr were found in V1500 Cyg (Harrison \& Campbell 2016) and V1432 Aql (Staubert et al. 2003), respectively, suggesting 
that the cause was likely a nova explosion in the past. The lack of nova shells
around the 4 systems is not however a clue against the nova interpretation, since
many old novae do not show such relics either (Pagnotta \& Zurek 2016).

Therefore, given the closer similarities of spin-to-orbit period ratios  with Paloma and IGR\,J19552+0044, rather than to the four asynchronous polars, we suggest \rx\, is in a similar evolutionary stage. A spectro-polarimetric study will prove tight constraints on the magnetic field and geometry of this new interesting mCV and its evolution.

\begin{table}
\caption{Main properties of the slightly desynchronised mCVs currently known (in order of increasing degree of asynchronism).}
\label{tab:mcv_properties}
\centering
\begin{tabular}{@{}lcccc}
\hline
System 				& $P_{{\rm spin}}$ 	& $P_{{\rm orb}}$ 	& $P_{{\rm spin}}$/$P_{{\rm orb}}$	& $P_{{\rm beat}}$\\
					& (s)				& (s)				&							& (hr)		 \\
\hline	
V 1432 Aquilae 	 		& 12\,150			& 12\,116			& 1.002						& 1\,213.4		 \\		
CD Ind				& 6\,579			& 6\,649			& 0.989						& 175.2		 \\
BY Camelopardalis		& 11\,961			& 12\,089			& 0.989						& 348.5		 \\
\rx\					& 5\,340			& 5\,940			& 0.898						& 15.2		 \\
Paloma				& 7\,800			& 9\,360			& 0.833						& 13.0		 \\ 
IGR\,J19552$+$0044 	& 4\,960			& 6\,100			& 0.813						& 7.3			 \\
EX Hydrae			& 4\,022			& 5\,895			& 0.682						& 3.5			 \\
\hline
\end{tabular}  
\begin{list}{}{}
\item[$^{*}$] The listed values are taken from: V 1432 Aquilae: Mukai et al. (2003); CD Ind: Myers et al. (2017); BY Camelopardalis: Silber et al. (1997); \rx\ (this study); Paloma: Joshi et al. (2016); IGR\,J19552$+$0044: Bernardini et al. (2013); EX Hydrae: Mauche et al. (2009).    
\end{list}
\end{table}

\subsection{\tmsp: a candidate transitional millisecond pulsar?}

The identification of \rx\, as an asynchronous polar argues against its identification as a possible counterpart of the \fermi-LAT gamma-ray source \fermirx, as emission at $\sim$GeV energies has never been reported from CVs, so far. On possibility might be that the gamma-ray emission is associated to one of the two faint X-ray sources detected in the \fermi-LAT error circle. 

As reported in Tab.\,\ref{tab:log_sources}, the absorption column density derived from the X-ray spectrum of \tmsp\, is $N_{\rm H} \sim 3 \times 10^{20}$ cm$^{-2}$, about one order of magnitude lower than the total Galactic value in the direction of the source. Assuming that the full observed column density contributes to the optical/NIR absorption, we estimate a colour excess $E(B-V) = 0.05$ mag (using the $N_{\rm H}/E(B-V)$ conversion of Predehl \& Schmitt 1995). From the results of our optical and NIR photometry (Sect. 3.1 and 3.3) we obtain for this source a de-reddened $B-R$ colour in the range 1.5-2.5 mag (taking into account the source variability and the uncertainties in the photometry) and $H-K = 0.3 \pm 0.2$ mag, broadly consistent with a late-type K2-M5 main sequence star. Under this hypothesis, using the distance modulus, the constraints on the source distance are very broad, ranging from 0.7 (M5) to 6.2 kpc (K2). 

In light of the observed colours, a coronal flare from an M-type star could be an explanation for the brightening observed in the \xmm\, X-ray and optical light curves shown in Fig.~\ref{src2_flare}. However, considering the inferred distance, the average X-ray flux ($\sim 8.5 \times 10{-14}$ erg cm$^{-2}$ s$^{-1}$ in the 0.2-10 keV, excluding the flare) would translate into a luminosity $ 5 \times 10^{30} < L_X  < 4 \times 10^{32}$ erg s$^{-1}$, which is too high for an active M-type star in ``quiescence'' (see, e.g., Stelzer et al. 2013).

An interesting possibility is the association of \fermirx\  with the X-ray variable source \tmsp, as a possible transitional millisecond pulsar (Archibald et al. 2009; Papitto et al. 2013). Such sources are often found in a sub-luminous disc state characterized by enhanced X-ray variability between a high and a low flux level that differ by a factor of $\sim$10, and often show occasional flaring activity on several minutes timescales (de Martino et al. 2013; Bogdanov et al. 2015), as we observe from \tmsp. A relatively bright emission is also observed at gamma-ray (de Martino et al. 2010, Stappers et al. 2014; Li et al. 2014), optical (de Martino et al. 2013; Bogdanov et al. 2015) and radio (Hill et al. 2011; Deller et al. 2015) wavelengths. Even though the absence of an ATCA radio detection from \tmsp\ does not favour its identification as a transitional millisecond pulsar, the non detection of a radio source might be compatible with a larger distance of the putative pulsar. In this respect, we note that the gamma-ray flux of \fermirx\,  ($\sim$$10^{-9}$\,ph cm$^{-2}$ s$^{-1}$ in the 1--100 GeV band; Acero et al. 2015), and its average X-ray flux ($\sim$$8.5\times10^{-14}$\,erg cm$^{-2}$ s$^{-1}$ in the 0.2--10 keV band) are $\sim$10 and 100 times smaller, respectively, than those observed from from the prototypical transitional millisecond pulsar PSR J1023$+$0038, which is located at 1.4\,kpc (Deller et al. 2012), hence a larger distance would be also in line with other multi-band source characteristics. Furthermore, the low spectral index and low level of variability of the GeV emission are also inline with what observed for other binary millisecond pulsars observed by \fermi-LAT (Acero et al. 2015).

The detection of optical/NIR emission from a late-type companion star, as inferred from the observed colors, would suggest for a redback system observed in the radio pulsar state. The constraints on the X-ray luminosity reported above would be in agreement with such a scenario. However, without knowing the orbital period, the amount of possible irradiation effects, the companion spectral type and the system distance, all the above considerations should be considered with some caution and alternative scenarios (e.g. a black-widow system) could not be ruled out.

\section*{Acknowledgements} 
The scientific results reported in this article are based on observations obtained with \xmm\ and \swift. 
\xmm\ is an ESA science mission with instruments and contributions directly funded by ESA Member 
States and the National Aeronautics and Space Administration (NASA). \swift\ is a NASA mission with 
participation of the Italian Space Agency and the UK Space Agency. This research has made use of 
softwares and tools provided by the High Energy Astrophysics Science Archive Research Center 
(HEASARC), which is a service of the Astrophysics Science Division at NASA/GSFC and the High 
Energy Astrophysics Division of the Smithsonian Astrophysical Observatory. The research has also 
made use of data from the Two Micron All Sky Survey, which is a joint project of the University of 
Massachusetts and the Infrared Processing and Analysis Center/California Institute of Technology, 
funded by NASA and the National Science Foundation. We also made use of data products from the 
Wide-field Infrared Survey Explorer, which is a joint project of the University of California, Los Angeles, 
and the Jet Propulsion Laboratory/California Institute of Technology, funded by NASA. The CANICA project 
is funded by the CONACyT grant G28586-E (PI: Luis Carrasco). This research is based on observations 
collected at the European Organization for Astronomical Research in the Southern Hemisphere under ESO 
program 296.D-5034 (PI: Rea). The results reported in this article are also based on observations made by
the Australia Telescope Compact Array (ATCA), which is part of the Australia Telescope National Facility that
is funded by the Australian Government for operation as a National Facility managed by CSIRO. This article is based on observations made with the IAC80 operated on the island of Tenerife by the IAC in the Spanish Observatorio del Teide. We are grateful to the XMM-Newton, ESO and ATCA Directors for awarding 
us their proprietary Director's Time. We thank Elisa Costantini for useful discussion, and deeply acknowledge the referee for his/her useful comments and suggestions. NR, FCZ, PE and AB 
are supported by the NWO Vidi Grant A.2320.0076 and by the European COST Action MP1304 (NewCOMPSTAR). 
NR, FCZ, DFT, EdOW and JL are also supported by grants AYA2015-71042-P and SGR2014-1073. FCZ, DFT, AP, 
EB and CF acknowledge the International Space Science Institute (ISSI-Bern) which funded and hosted the 
international team "The disc-magnetosphere interaction around transitional millisecond pulsars". DdM 
acknowledges support from INAF--ASI I/037/12/0. AP acknowledges support via an EU Marie Sk\l odowska 
Curie fellowship under grant no. 660657-TMSP-H2020-MSCA-IF-2014. MH is supported by grants ESP2015-66134-R, 
SGR2014-1458 and FEDER funds. ML and JC acknowledge support by the Spanish grant AYA2013-42627. M.L. is also supported by EU's Horizon 2020 programme through a Marie Sklodowska-Curie Fellowship (grant nr. 702638). JC also acknowledges support by the Leverhulme Trust through the Visiting Professorship Grant VP2-2015-046.

\bsp

\section{Appendix}

\begin{figure}
\centering
\resizebox{\hsize}{!}{\includegraphics{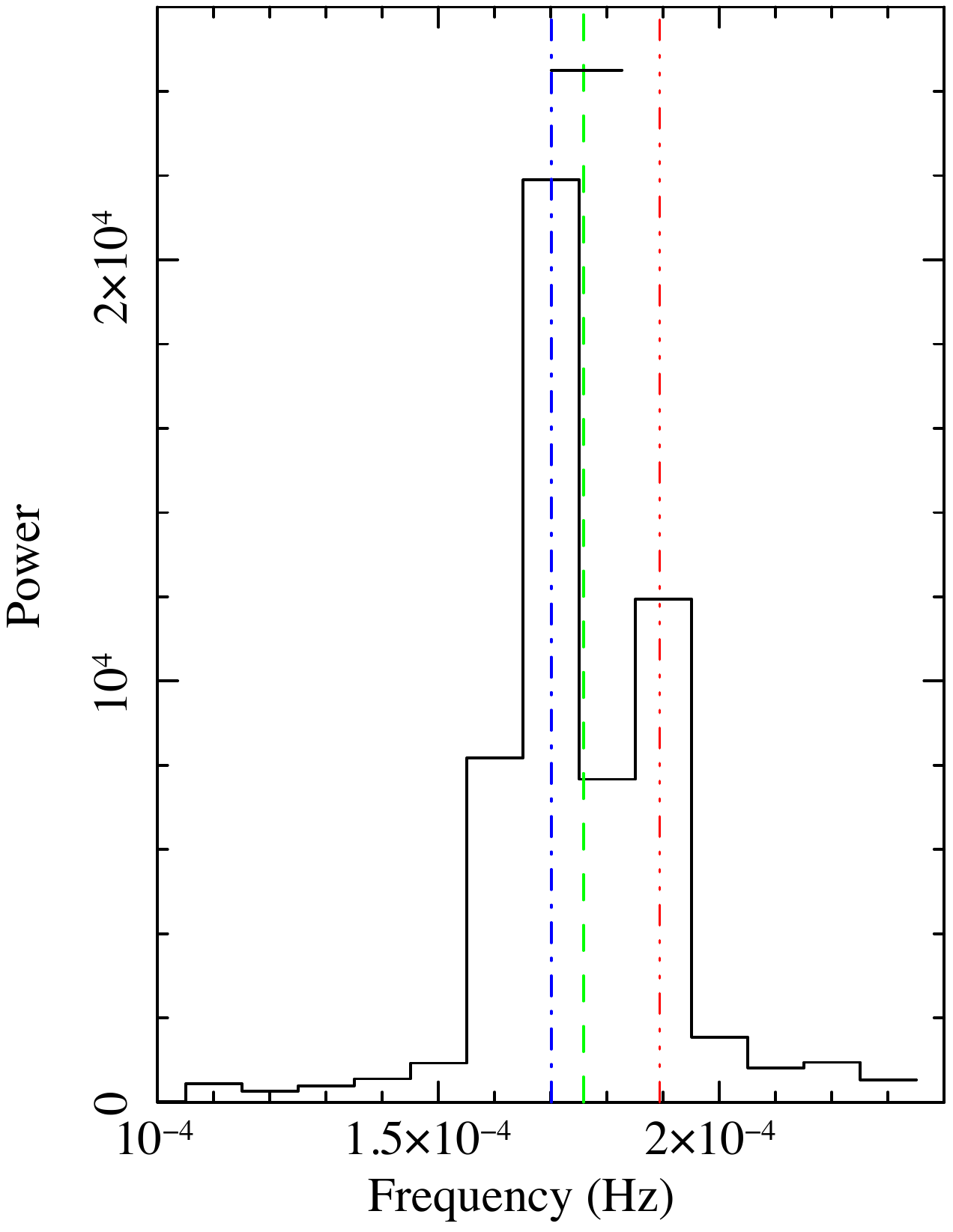}\includegraphics{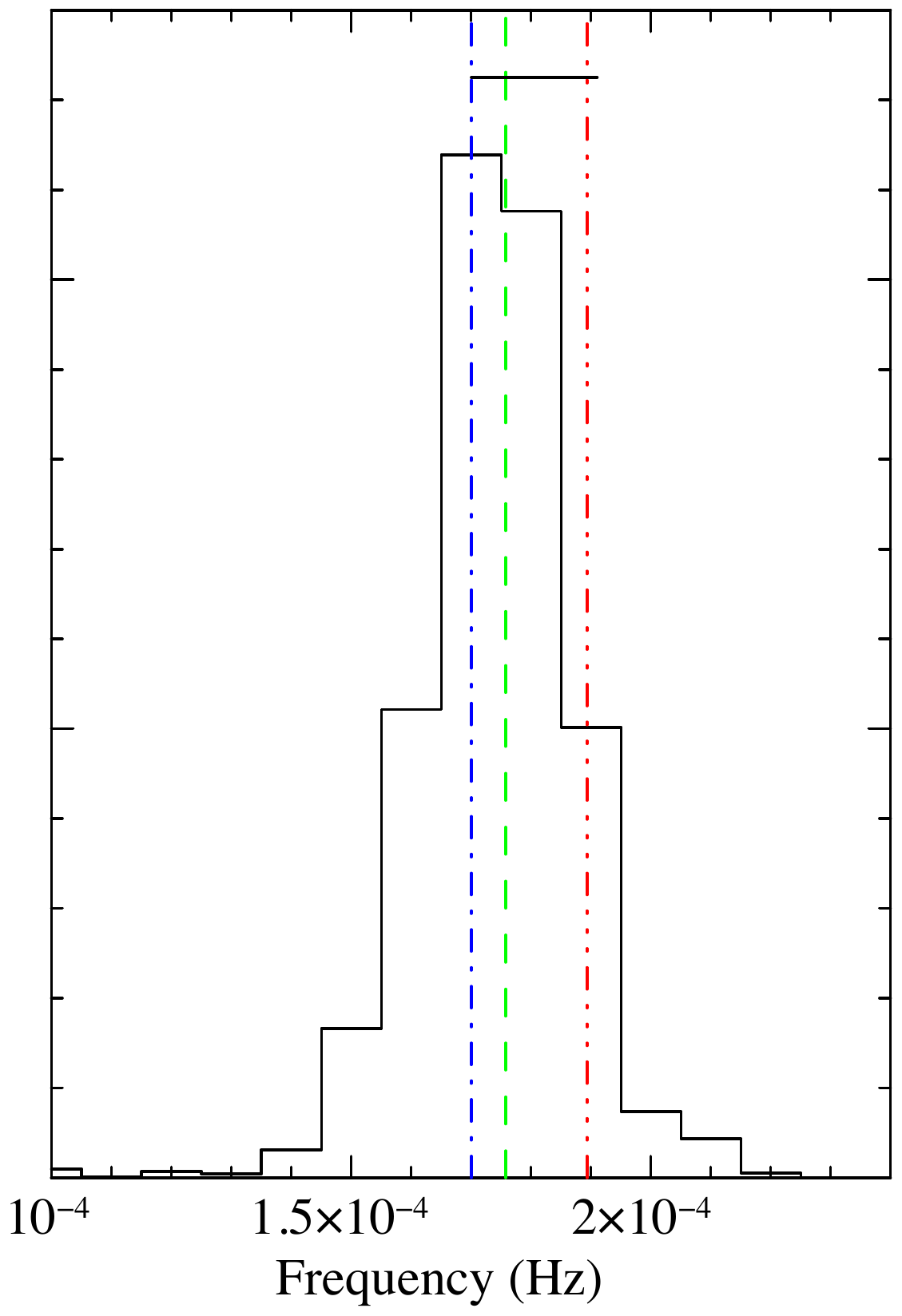}}
\caption{\label{peak_compare} Zoom on the PSD structure around $\Omega$ (blue dot-dash line) and $\omega$ (red 3dot-dash line) in the second \xmm\ observation (left) and only in the first 45 ks of it (before the flux minimum and the possible phase jump, see Sect.\,\ref{xmmtiming}). The dashed green line indicates the frequency of the signal reported by Halpern et al. (2017). The length of the horizontal bar at the top indicates the Fourier intrinsic resolution of the data segment.}
\end{figure}

During the referee process of this paper, another work using our \xmm\, data were published by Halpern, Bogdanov \& Thorstensen (2017; ApJ, 838, 124).  In this Appendix we point out a few differences between the two publications. \\
1) In our optical monitoring we could not detect any hint for periodicity in the optical monitoring of the transitional MSP candidate \tmsp, nor the broad minimum reported by Halpern et al. (2017), but we caution, that the longest time-series we collected span over about 2.9 hours (about one half with respect to the length of Halpern et al. optical dataset).\\
2) By using our \xmm\ data, they report different modelling/interpretation of the X-ray light curve. They assumed that a phase jump of $\sim$120\deg\ in the $\sim$1.6\,h  modulation happens around the flux minimum and, using only the first part of the observations, derived a spin period of $\approx$1.58~h, which is barely compatible with our value. As for the orbital period, they measured it from radial velocities of H$\alpha$ in good quality optical spectroscopic data and found $1.64\pm0.06$~h, in good agreement with our values. These differences reflects in a different interpretation on the nature of the mCV \rx. To see whether one description of the light curve is to be preferred to the other, we analysed separately the \xmm\ segments of data prior and after their suggested phase jump in both observations. The comparison was inconclusive, essentially because the orbital and spin periods derived in the two scenarios are almost identical and because not all subsets of data have enough frequency resolution to separate the $\Omega$ and $\omega$ peaks in the PSDs (see Fig.\,\ref{peak_compare} for an example).

\label{lastpage}
\end{document}